\documentclass[12pt, twocolumn, trackchanges]{aastex701}

\usepackage{amssymb}
\usepackage{amsmath} 

\usepackage{epstopdf}
\usepackage{color}
\usepackage{amsmath}
\usepackage{graphicx}
\usepackage{hyperref}
\usepackage{graphics,graphicx,natbib}
\usepackage{multirow}
\usepackage{xspace}
\usepackage{makecell}

\def\lapp{\ifmmode\stackrel{<}{_{\sim}}\else$\stackrel{<}{_{\sim}}$\fi}
\def\gapp{\ifmmode\stackrel{>}{_{\sim}}\else$\stackrel{>}{_{\sim}}$\fi}

\begin{document}
\title{FRB 20210601A: Extreme Temporal Clustering in a Repeating Fast Radio Burst Source as Evidence for a Hidden Class of Episodic Repeaters}
\shorttitle{FRB 20210601A: Extreme Temporal Clustering}
\shortauthors{R. Mckinven, \emph{et al.}}

\correspondingauthor{Ryan Mckinven}
\email{ryan.mckinven@mcgill.ca}

\author[0000-0001-7348-6900]{Ryan Mckinven}
\email{ryan.mckinven@mcgill.ca}
\affiliation{Department of Physics, McGill University, 3600 rue University, Montr\'eal, QC H3A 2T8, Canada}
\affiliation{Trottier Space Institute, McGill University, 3550 rue University, Montr\'eal, QC H3A 2A7, Canada}


\author[0000-0001-6422-8125]{Amanda M. Cook}
\email{amanda.cook@mail.mcgill.ca}
\affiliation{Department of Physics, McGill University, 3600 rue University, Montr\'eal, QC H3A 2T8, Canada}
\affiliation{Trottier Space Institute, McGill University, 3550 rue University, Montr\'eal, QC H3A 2A7, Canada}
\affiliation{Anton Pannekoek Institute for Astronomy, University of Amsterdam, Science Park 904, 1098 XH Amsterdam, The Netherlands}

\author[0000-0002-8376-1563]{Alice P. Curtin}
\email{alice.curtin@mail.mcgill.ca}
\affiliation{Department of Physics, McGill University, 3600 rue University, Montr\'eal, QC H3A 2T8, Canada}
\affiliation{Trottier Space Institute, McGill University, 3550 rue University, Montr\'eal, QC H3A 2A7, Canada}
\affiliation{Anton Pannekoek Institute for Astronomy, University of Amsterdam, Science Park 904, 1098 XH Amsterdam, The Netherlands}

\author[0000-0002-3382-9558]{B. M. Gaensler}
\email{gaensler@ucsc.edu}
\affiliation{Department of Astronomy and Astrophysics, University of California, Santa Cruz, 1156 High Street, Santa Cruz, CA 95060, USA}
\affiliation{Dunlap Institute for Astronomy and Astrophysics, 50 St. George Street, University of Toronto, ON M5S 3H4, Canada}
\affiliation{David A.\ Dunlap Department of Astronomy and Astrophysics, 50 St. George Street, University of Toronto, ON M5S 3H4, Canada}

\author[0000-0003-2317-1446]{Jason W.T. Hessels}
\email{jason.hessels@mcgill.ca}
\affiliation{Department of Physics, McGill University, 3600 rue University, Montr\'eal, QC H3A 2T8, Canada}
\affiliation{Trottier Space Institute, McGill University, 3550 rue University, Montr\'eal, QC H3A 2A7, Canada}
\affiliation{Anton Pannekoek Institute for Astronomy, University of Amsterdam, Science Park 904, 1098 XH Amsterdam, The Netherlands}
\affiliation{ASTRON, Netherlands Institute for Radio Astronomy, Oude Hoogeveensedijk 4, 7991 PD Dwingeloo, The Netherlands}

\author[0009-0009-0938-1595]{Naman Jain}
\email{naman.jain@mail.mcgill.ca}
\affiliation{Department of Physics, McGill University, 3600 rue University, Montr\'eal, QC H3A 2T8, Canada}
\affiliation{Trottier Space Institute, McGill University, 3550 rue University, Montr\'eal, QC H3A 2A7, Canada}

\author[0009-0007-5296-4046	]{Lordrick Kahinga}
\email{lkahinga@ucsc.edu}
\affiliation{Department of Astronomy and Astrophysics, University of California, Santa Cruz, 1156 High Street, Santa Cruz, CA 95060, USA}
\affiliation{Department of Physics, College of Natural and Mathematical Sciences, University of Dodoma, 1 Benjamin Mkapa Road, 41218 Iyumbu, Dodoma, Tanzania}

\author[0000-0001-9345-0307]{Victoria M. Kaspi}
\email{victoria.kaspi@mcgill.ca}
\affiliation{Department of Physics, McGill University, 3600 rue University, Montr\'eal, QC H3A 2T8, Canada}
\affiliation{Trottier Space Institute, McGill University, 3550 rue University, Montr\'eal, QC H3A 2A7, Canada}
\affiliation{School of Physics and Astronomy, Tel Aviv University, Tel Aviv 69978, Israel}

\author[0000-0002-1800-8233]{Charanjot Brar}
\email{charanjot.brar@nrc-cnrc.gc.ca}
    \affiliation{National Research Council of Canada, Herzberg Astronomy and Astrophysics, 5071 West Saanich Road, Victoria, BC V9E 2E7, Canada}

\author[0000-0003-2047-5276]{Tomas Cassanelli}
\email{tcassanelli@ing.uchile.cl}
\affiliation{Department of Electrical Engineering, Universidad de Chile, Av. Tupper 2007, Santiago 8370451, Chile}

\author[0000-0002-2878-1502]{Shami Chatterjee}
  \affiliation{Cornell Center for Astrophysics and Planetary Science, Cornell University, Ithaca, NY 14853, USA}
  \email{shami.chatterjee@cornell.edu}

\author[0000-0003-4098-5222]{Fengqiu Adam Dong}
\email{fadong@yorku.ca}
\affiliation{Department of Physics and Astronomy, York University, 4700 Keele Street, Toronto, ON MJ3 1P3, Canada}

\author[0009-0006-1258-4228]{Maxwell Fine}
\email{maxwell.fine@mail.mcgill.ca}
\affiliation{Department of Physics, McGill University, 3600 rue University, Montr\'eal, QC H3A 2T8, Canada}
\affiliation{Trottier Space Institute, McGill University, 3550 rue University, Montr\'eal, QC H3A 2A7, Canada}

\author[0000-0001-8384-5049]{Emmanuel Fonseca}
\email{emmanuel.fonseca@mail.wvu.edu}
\affiliation{Department of Physics and Astronomy, West Virginia University, PO Box 6315, Morgantown, WV 26506, USA}
\affiliation{Center for Gravitational Waves and Cosmology, West Virginia University, Chestnut Ridge Research Building, Morgantown, WV 26505, USA}

\author[0009-0004-4176-0062]{Afrokk Khan}
\email{afrasiyab.khan@mcgill.ca}
\affiliation{Department of Physics, McGill University, 3600 rue University, Montr\'eal, QC H3A 2T8, Canada}
\affiliation{Trottier Space Institute, McGill University, 3550 rue University, Montr\'eal, QC H3A 2A7, Canada}

\author[0009-0008-6166-1095]{Bikash Kharel}
  \email{bk0055@mix.wvu.edu}
  \affiliation{Department of Physics and Astronomy, West Virginia University, PO Box 6315, Morgantown, WV 26506, USA}
  \affiliation{Center for Gravitational Waves and Cosmology, West Virginia University, Chestnut Ridge Research Building, Morgantown, WV 26505, USA}

\author[0000-0002-4209-7408]{Calvin Leung}
\email{calvin_leung@berkeley.edu}
\affiliation{Miller Institute for Basic Research, Stanley Hall, Room 206B, Berkeley, CA 94720, USA}
 \affiliation{Department of Astronomy, University of California, Berkeley, CA 94720-3411, USA}

\author[0000-0002-7164-9507]{Robert Main}
\email{robert.main@mcgill.ca}
\affiliation{Department of Physics, McGill University, 3600 rue University, Montr\'eal, QC H3A 2T8, Canada}
\affiliation{Trottier Space Institute, McGill University, 3550 rue University, Montr\'eal, QC H3A 2A7, Canada}
\affiliation{Dominion Radio Astrophysical Observatory, Herzberg Astronomy \& Astrophysics Research Centre, National Research Council Canada, 717 White Lake Rd., Kaleden, BC, V0H 1K0 Canada}

\author[0000-0002-4279-6946]{Kiyoshi W. Masui}
\email{kmasui@mit.edu}
\affiliation{MIT Kavli Institute for Astrophysics and Space Research, Massachusetts Institute of Technology, 77 Massachusetts Ave, Cambridge, MA 02139, USA}
\affiliation{Department of Physics, Massachusetts Institute of Technology, 77 Massachusetts Ave, Cambridge, MA 02139, USA}

\author[0000-0003-2111-3437]{Kyle McGregor}
\email{kyle.mcgregor@mail.mcgill.ca}
\affiliation{Department of Physics, McGill University, 3600 rue University, Montr\'eal, QC H3A 2T8, Canada}
\affiliation{Trottier Space Institute, McGill University, 3550 rue University, Montr\'eal, QC H3A 2A7, Canada}


\author[0000-0001-7491-046X]{Nicole Mulyk}
\email{nicole.mulyk@mail.mcgill.ca}
\affiliation{Department of Physics, McGill University, 3600 rue University, Montr\'eal, QC H3A 2T8, Canada}
\affiliation{Trottier Space Institute, McGill University, 3550 rue University, Montr\'eal, QC H3A 2A7, Canada}

\author[0000-0002-0940-6563]{Mason Ng}
\email{mason.ng@mcgill.ca}
\affiliation{Department of Physics, McGill University, 3600 rue University, Montr\'eal, QC H3A 2T8, Canada}
\affiliation{Trottier Space Institute, McGill University, 3550 rue University, Montr\'eal, QC H3A 2A7, Canada}

\author[0000-0003-0510-0740]{Kenzie Nimmo}
\email{knimmo@northwestern.edu}
\altaffiliation{NASA Einstein Fellow.}
\affiliation{Center for Astrophysics $|$ Harvard \& Smithsonian, 60 Garden St, Cambridge, MA 02138, USA}

\author[0000-0002-8897-1973]{Ayush Pandhi}
\email{ayush.pandhi@mcgill.ca}
\affiliation{Department of Physics, McGill University, 3600 rue University, Montr\'eal, QC H3A 2T8, Canada}
\affiliation{Trottier Space Institute, McGill University, 3550 rue University, Montr\'eal, QC H3A 2A7, Canada}

\author[0000-0002-8912-0732]{Aaron~B.~Pearlman}
\email{aaron.b.pearlman@mit.edu}
\altaffiliation{NASA Hubble Fellow.}
\affiliation{MIT Kavli Institute for Astrophysics and Space Research, Massachusetts Institute of Technology, 77 Massachusetts Ave, Cambridge, MA 02139, USA}
\affiliation{Department of Physics, McGill University, 3600 rue University, Montr\'eal, QC H3A 2T8, Canada}
\affiliation{Trottier Space Institute, McGill University, 3550 rue University, Montr\'eal, QC H3A 2A7, Canada}

\author[0000-0002-4795-697X]{Ziggy Pleunis}
\email{z.pleunis@uva.nl}
\affiliation{Anton Pannekoek Institute for Astronomy, University of Amsterdam, Science Park 904, 1098 XH Amsterdam, The Netherlands}
\affiliation{ASTRON, Netherlands Institute for Radio Astronomy, Oude Hoogeveensedijk 4, 7991 PD Dwingeloo, The Netherlands}

\author[0000-0002-4623-5329]{Mawson W. Sammons}
\email{mawson.sammons@mcgill.ca}
\affiliation{Department of Physics, McGill University, 3600 rue University, Montr\'eal, QC H3A 2T8, Canada}
\affiliation{Trottier Space Institute, McGill University, 3550 rue University, Montr\'eal, QC H3A 2A7, Canada}

\author[0000-0002-7374-7119]{Paul Scholz}
\email{pscholz@yorku.ca}
\affiliation{Department of Physics and Astronomy, York University, 4700 Keele Street, Toronto, ON MJ3 1P3, Canada}

\author[0000-0002-6823-2073]{Kaitlyn Shin}
\email{kaitshin@caltech.edu}
\affiliation{Cahill Center for Astronomy and Astrophysics, MC 249-17 California Institute of Technology, Pasadena CA 91125, USA}

\author[0000-0003-2631-6217]{Seth R. Siegel}
\email{Seth.Siegel@skao.int}
\affiliation{SKAO, Science Operations Centre, CSIRO ARRC, 26 Dick Perry Avenue, Kensington WA 6151 Australia}
\affiliation{Perimeter Institute of Theoretical Physics, 31 Caroline Street North, Waterloo, ON N2L 2Y5, Canada}
\affiliation{Department of Physics, McGill University, 3600 rue University, Montr\'eal, QC H3A 2T8, Canada}
\affiliation{Trottier Space Institute, McGill University, 3550 rue University, Montr\'eal, QC H3A 2A7, Canada}

\author[0000-0002-2088-3125]{Kendrick Smith}
\email{kmsmith@perimeterinstitute.ca}
\affiliation{Perimeter Institute of Theoretical Physics, 31 Caroline Street North, Waterloo, ON N2L 2Y5, Canada}

\author[0000-0001-9784-8670]{Ingrid Stairs}
  \email{stairs@astro.ubc.ca}
  \affiliation{Department of Physics and Astronomy, University of British Columbia, 6224 Agricultural Road, Vancouver, BC V6T 1Z1 Canada}


\begin{abstract}
  
Population studies of fast radio burst (FRB) repetition typically model sources using a single distribution of steady, time-averaged burst rates, yet individual sources show significant departures from this assumption. Here we present FRB 20210601A, a source discovered by the Canadian Hydrogen Intensity Mapping Experiment FRB project (CHIME/FRB) that represents the most extreme such departure yet identified. The source produced at least nine bursts within a single 106.5~s interval -- the highest instantaneous burst rate yet reported from a repeating FRB source at $\gtrsim 2$~Jy~ms fluences -- yet remained otherwise undetected despite $\sim 320$~hr of exposure spanning 7~yr. The active interval occurred while the source was in a side-lobe of the CHIME/FRB digitized beams, implying an instantaneous burst rate $\gtrsim2.3\rm{\times}$ higher ($\gtrsim$ 21 bursts) at main-lobe sensitivity. The source's intense, brief burst storm flanked by apparent quiescence despite substantial exposure is unique among the CHIME/FRB repeater sample, although its burst morphology, excess dispersion measure ($\rm{DM_{excess}} \sim 360\;\rm{pc\,cm^{-3}}$), Faraday rotation measure ($|\rm{RM_{excess}}| \sim 150-170\;\rm{rad\,m^{-2}}$), and temporal scattering ($\tau \lesssim 0.37$ ms at 600 MHz) are otherwise typical. Modeling a population of episodically active repeaters (i.e., short burst storms separated by long recurrence times) and accounting for declination-dependent exposure, we forecast $\sim1-4$ additional source discoveries similar to FRB~20210601A over the next decade of CHIME/FRB operations, corresponding to a CHIME/FRB-detectable population of $9^{+200}_{-8}$ sources (68\% credible interval). These results demonstrate that highly episodic repeaters could constitute a selection-biased but detectable component of the FRB population, possibly reflecting distinct magnetospheric states or triggering mechanisms.
\end{abstract}

\keywords{Radio transient sources (2008), Compact objects (288)}

\section{Introduction}
\label{sec:intro}

The origin of fast radio bursts \citep[FRBs;][]{Lorimer2007,Petroff2022} is closely tied to a central question: do all FRBs repeat \citep[e.g.,][]{Spitler2016,Palaniswamy2018}? This question is motivated by the possibility of multiple progenitor channels, including cataclysmic scenarios that would preclude repetition for a subset of the population \citep[e.g.,][]{Totani2013,Falcke2014,Zhang2023}. Despite a steadily growing sample of thousands of FRB detections and rapid progress in localizing and characterizing repeating FRBs, the relationship between repeaters and one-off events remains unsettled, with different observational properties pointing toward competing interpretations \citep[e.g.,][]{Pleunis2021, Gordon2023, James2023}. Repeating sources provide a uniquely informative window into FRB physics, enabling targeted follow-up and long-term monitoring that have revealed key phenomenology including sub-burst frequency drifts \citep[e.g.,][]{Hessels2019}, extreme and time-variable Faraday rotation measures \citep[e.g.,][]{michilli2018extreme}, and diverse host galaxy environments \citep[e.g.,][]{Chatterjee2017,Tendulkar2017,Marcote2020}. 
However, the vast majority of detected FRBs have not been observed to repeat \citep{abb+23, Cook2026}, and whether repeaters and apparent one-offs represent distinct source classes or different manifestations of a single underlying population remains a central question in the field.

The Canadian Hydrogen Intensity Mapping Experiment Fast Radio Burst Project (CHIME/FRB) \citep{2018ApJ...863...48C} has transformed the statistical study of repetition by leveraging its wide field-of-view and daily, uniform monitoring of the northern sky ($\rm{Dec}\gtrsim -10^{\circ}$) over $400-800$ MHz. Population analyses of the first CHIME/FRB catalog \citep{aab+21} and subsequent studies of the repeating subsample \citep{abb+23} demonstrated that the observed fraction of repeaters and apparent one-offs can be reproduced by a single underlying distribution of repeat rates \citep{James2023,McGregor2024,Yamasaki2024}. In the second CHIME/FRB catalog \citep{CHIME/FRB2026a}, consisting of 3641 unique sources, only $\sim2.4\pm0.4$\% of sources are observed to repeat. Nevertheless, the lowest measured repetition rates smoothly connect to the upper limits for one-off sources, consistent with a unified population interpretation \citep{Cook2026}\footnote{This need not preclude diversity along other dimensions that may still distinguish a minority sub-population (e.g., host environment, persistent radio source association, etc.).}.

These repetition analyses implicitly assume that burst arrival times can be described by stationary stochastic processes (i.e., processes with constant statistical properties over time), typically Poissonian or Weibull-distributed, characterized by a single time-averaged rate \citep[e.g.,][]{James2023,McGregor2024,Yamasaki2024}. Within this framework, apparent non-repeaters are interpreted as intrinsically weak or relatively inactive repeating sources whose additional bursts fall below current detection thresholds or outside the nominal exposure window of the survey \citep[e.g.,][]{Good2023}. Detailed studies of individual repeaters, however, increasingly demonstrate departures from stationarity including clustered burst activity or ``burst storms" \citep[e.g.,][]{Nimmo2023}, evolving repetition rates, bi-modal waiting-time distributions \citep[e.g.,][]{2018MNRAS.475.5109O, Li2021, Cruces2021, 2021ApJ...920L..23Z,2022RAA....22l4002Z, Wang2025}, abrupt changes in activity \citep[e.g., FRB~20201124A;][]{Lanman2022,Xu2022}, and periodic or quasi-periodic activity windows \citep[e.g., FRB~20180916B, FRB~20121102A;][]{aab+20,Rajwade2020, Cruces2021,Sand2023}. A broader search for periodicity across the CHIME/FRB repeater sample is also underway (Pearlman et al. 2026, in prep.). Strongly episodic activity confined to brief, rare intervals may therefore be underrepresented in time-averaged population statistics. 

In this context, the study of statistical outliers complements population modeling by probing repetition regimes that may be rare, transient, or strongly selection biased. FRB~20210601A, originally published in the second CHIME/FRB Catalog \citep{CHIME/FRB2026a}, is an extreme case within the CHIME/FRB sample. As we show here, all nine detected bursts occurred within a single interval spanning less than 2 minutes, followed and preceded by years of continued monitoring without recurrence. 
This work presents a detailed analysis of FRB~20210601A, including its spectro-temporal and polarization properties, repetition behavior, localization, and counterpart constraints. We show that its burst statistics are not only inconsistent with stationary repetition models, but place it as an extreme outlier within the CHIME/FRB repeater sample. We argue instead for a population of highly episodic repeaters, characterized by short burst storms separated by long quiescent intervals, and we develop a simplified population framework that incorporates the declination-dependent exposure of CHIME/FRB to constrain this scenario and to forecast future detections.

\section{Observations}
\label{sec:obs}

On 2021 June 1, CHIME/FRB detected a series of nine bursts from FRB 20210601A within 106.5 s, all above the real-time detection threshold of S/N$\geq 8.5$. Throughout this work, we define a burst as a single real-time intensity trigger, regardless of internal sub-structure (e.g., multiple sub-components). In effect, this counts multi-component bursts with separations of $\lesssim 100$~ms as the same burst, set by the maximum duration of a single intensity data dump; components separated by longer intervals would be captured as independent triggers. This assumption is made for simplicity and is in keeping with convention adopted when reporting CHIME/FRB sources, repeating or otherwise \citep{CHIME/FRB2026a}.

Total intensity data ingested by the real-time system were successfully recorded for all nine events, with two of the events also successfully captured as raw voltage (baseband) data \citep{2018ApJ...863...48C}. Due to a contemporaneous network issue affecting event registration, these bursts did not trigger real-time CHIME/FRB VOEvent alerts \citep{2025AJ....169...39A} and were identified later during a systematic database audit in December 2024 during the construction of the second CHIME/FRB Catalog \citep{CHIME/FRB2026a}. Of these nine bursts, eight (FRB~20210601A, FRB~20210601C, FRB~20210601D, FRB~20210601E, FRB~20210601F, FRB~20210601G, FRB~20210601[H,I]\footnote{FRB~20210601H and FRB~20210601I were found to be duplicate events upon subsequent inspection owing to detection across multiple tree indices of CHIME/FRB's {\tt Bonsai} search pipeline \citep{2018ApJ...863...48C}.}, FRB~20210601J) were reported by \citet{CHIME/FRB2026a}; one burst, FRB~20210601B, was excluded owing to its low S/N and a tentative ``FAINT/NOISE'' classification. Subsequent inspection of the data, however, reveals signal well above the noise floor despite its modest S/N, and its occurrence during the active interval lends high confidence that it is a genuine detection.

The nine bursts share a roughly consistent dispersion measure of $\mathrm{DM} \sim 419\;\rm{pc\,cm^{-3}}$. Galactic free electron density models at the position of the source ($\ell, b \sim 117.79^\circ, 50.97^\circ$; see Section~\ref{sec:localization}) predict modest DM contributions from the Milky Way: 
$\sim 32\; \rm{pc\, cm^{-3}}$ \citep[NE2025;][]{Ocker2026}
and $26\; \rm{pc\, cm^{-3}}$ \citep[YMW16;][]{ymw17}, respectively. Assuming a Galactic halo contribution of $\mathrm{DM}_{\rm{halo}}\sim30\; \rm{pc\, cm^{-3}}$ \citep{Dolag2015, Cook2023}, we infer $\mathrm{DM}_{\rm{excess}}\sim 360\; \rm{pc\, cm^{-3}}$ for the source.

\begin{figure*}
	\centering
    \includegraphics[width=0.8\textwidth]{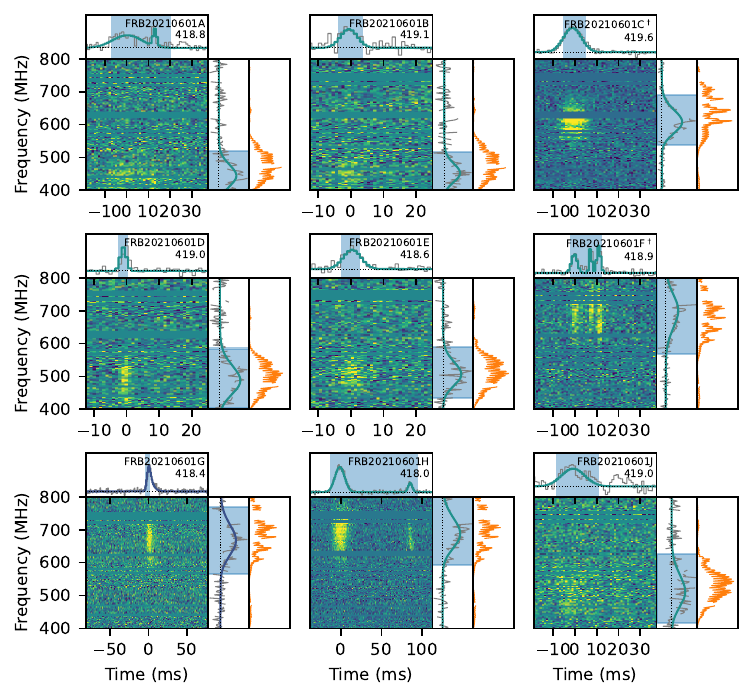}
    \caption{A reproduction of the de-dispersed dynamic spectra (``waterfall" plots) of the 8 detections of FRB~20210601A originally reported in CHIME/FRB Catalog 2 \citep{CHIME/FRB2026a}, plus an additional detection (FRB 20210601B). Bursts are ordered by arrival time with the TNS name and best-fit dispersion measure (DM, in pc cm$^{-3}$) displayed in the upper right corner of each panel. Frequency and time axes are displayed at 3.125 MHz and 0.983 ms resolutions, respectively. Blue shaded regions mark the burst duration (time series) and apparent emission bandwidth (spectrum); frequency and time averaged profiles use only data within these regions and are accompanied by model fits \citep[See][for details]{CHIME/FRB2026a}. The predicted beam response is displayed in a separate panel in orange for each event. Events appended with $\dagger$ symbol (FRB 20210601C and FRB 20210601F) indicate the existence of higher temporal resolution waterfalls plots produced from baseband data shown in Figure~\ref{fig:baseband_waterfall}.}
    \label{fig:waterfalls}
\end{figure*}

Figure~\ref{fig:waterfalls} shows the dynamic spectra (``waterfall" plots) produced from the intensity data recorded for the nine events. 
All nine bursts from FRB~20210601A were detected off meridian, in the first side-lobe of CHIME/FRB's digital search beams, just prior to transit through CHIME's main sensitivity lobe. The spectral properties of each detected event are found to closely match the observed chromaticity in sensitivity predicted by CHIME's beam model\footnote{\url{https://chime-frb-open-data.github.io/beam-model/}} (see Figure~\ref{fig:beam_map}), strongly suggesting that the observed spectral properties are dominated by the instrument response rather than are intrinsic to the source. Indeed, an adjoining panel for each burst in Figure~\ref{fig:waterfalls} shows the predicted beam response across CHIME's $400-800$ MHz bandpass, which broadly tracks the observed emitting band. This reinforces that the observed spectral properties are set by the instrumental beam response and that the intrinsic spectrum of each burst is likely to be significantly broader. 

A systematic search for additional repeat bursts was extended through 31 December 2025, querying our event database for any candidate repeat bursts at similar sky position and with $\mathrm{S/N} \geq 8.5$ and $413.2 \leq \rm{DM} \leq 425.2 \; \rm{pc\, cm}^{-3}$. This search included events detected off-axis relative to the CHIME/FRB search beams, motivated by the possibility that bright bursts could be detected through side-lobes and therefore not be automatically associated with FRB 20210601A by the real-time pipeline or the offline DBSCAN-based clustering algorithm  
\citep[see][for details]{Cook2026,Cook2026B}. No additional bursts were identified from this search. We also searched for an increased incidence of sub-threshold events with $7.5 \leq \mathrm{S/N} < 8.5$, which are recorded in the event database but do not meet the threshold required to trigger callback data capture. We find no evidence for an elevated rate of such sub-threshold events in the days immediately before or after 1 June 2021, disfavoring the scenario in which the activity window extended beyond the June 1 transit but at levels just below the detection threshold.


Exposure calculations for this source estimate a cumulated exposure of $\sim 220$ hours over the Catalog 2 time frame \citep{Cook2026}. Updating our exposure calculation to the end of December 2025 yields $\sim 320$ hours of total exposure; $\sim 132$ hours until first detection and $\sim 188$ hours since the last detection. We note that this exposure is not continuous but is accumulated in $\sim 10$~min snapshots once per sidereal day as the source transits CHIME, corresponding to 2025 individual transits with non-zero exposure between 25 July 2018 and 31 December 2025, inclusive (i.e., excluding days when the instrument was not operational). Above a $95\%-$fluence completeness threshold of $\sim 7$~Jy~ms, we infer a burst rate of $R = 0.028^{+0.025}_{-0.015}$ bursts~hr$^{-1}$ over the full exposure window (95\% confidence; 9 bursts in 320~hr), an upper limit of $R \lesssim 0.023$ bursts~hr$^{-1}$ over the interval preceding first detection (95\% confidence; 0 bursts in 132~hr), and an upper limit of $R \lesssim 0.016$ bursts~hr$^{-1}$ over the apparently quiescent interval following the last detection (95\% confidence; 0 bursts in 188~hr).\footnote{For simplicity, we assume Poissonian statistics here, which provides a standard benchmark rate for comparison with other sources.}
We note that since the nine bursts were detected in the side-lobes of CHIME/FRB's digitized search beams, all nine events fell outside the nominal per-day exposure window, and the exposure estimate does not account for side-lobe sensitivity. Thus, within the context of setting an upper limit on the burst rate during the quiescent interval, the estimated exposure used is a lower limit, implying the above rate upper limit is an overestimate. The implications of the discrete, narrow sampling window of CHIME/FRB on apparent source activity are explored further in Section~\ref{discussion:modelling}.

The average daily exposure and sensitivity to FRB 20210601A have declined since its initial detection due to the steady drift of the source through the CHIME primary beam, arising from Earth’s precession. At the time of detection, the daily exposure was approximately 10 minutes. This has gradually declined at a rate of roughly 0.3 minutes per year, corresponding to an annual reduction of about 3\% in exposure. Using the CHIME/FRB beam model\footnote{\url{https://chime-frb-open-data.github.io/beam-model/}}, we estimate a reduction in sensitivity at 400 and 800 MHz relative to the detection epoch of approximately 5\% and 20\%, respectively. The combined daily exposure and sensitivity loss from this gradual beam drift is comparable to that experienced by other CHIME/FRB sources, many of which nonetheless exhibit repeat bursts with waiting times of days to weeks between active transits. As such, the absence of subsequent bursts since 1 June 2021 cannot be attributed to this gradual decline in either exposure or sensitivity.

\section{Results}
\label{sec:results}

\subsection{Sky Position}
\label{sec:localization}

The baseband data collected for FRB~20210601C and FRB~20210601F were passed through the CHIME/FRB baseband localization pipeline \citep[see][for details]{Michilli2021} to obtain individual source localizations (R.A., Dec.; J2000, with $1\sigma$ uncertainties in parentheses) of $200.76(2)^{\circ}, +65.83(1)^{\circ}$ and $200.80(2)^{\circ}, +65.843(9)^{\circ}$, respectively.
We combine these individual measurements, accounting for both statistical and systematic uncertainty following the methodology outlined in Baseband Catalog 2 (CHIME/FRB Collaboration et al. 2026b, in prep.), to obtain a final best measured position of $200.780(14)^{\circ}, +65.837(7)^{\circ}$.

Figure~\ref{fig:localization} shows the Pan-STARRS1 3$\pi$ $i_{\rm P1}$-band stacked image \citep{chambers2016pan, Waters16} of the approximate localization region of FRB 20210601A, centered on the refined localization. To determine the association probability of any galaxies in the vicinity of FRB 20210601A, we utilize PATH \citep[Probabilistic Association of Transients to their Hosts;][]{Aggarwal2021}, a Bayesian framework for transient host association. PATH was applied to the DESI Legacy Imaging Surveys DR9 catalog \citep{Dey2019}, accessed through the NOIRLab Astro Data Lab \citep{Nikutta2020}. At this declination, the DR9 photometry comes from the BASS and MzLS surveys \citep{Zhou2017}. We specify the prior probability that the host is unseen/undetected to be 15\% (i.e., $P(U)=0.15$) to account for the limiting magnitude of the DR9 north data and an exponential offset prior with a scale length of 0.5, corresponding to half the galaxy's effective radius. Both the revised offset prior and the scale length adopted here are consistent with recent FRB host-association work \citep[e.g.,][]{Dong2025}. The highest-probability candidate host (Legacy Surveys DR9  ID 9907740743173799, $m_r \approx 20.2$; $z_{\rm phot} = 0.28 \pm 0.06$) located 8.5'' from the localization centroid, has $P(O|x) \approx 0.14$, comparable to the unseen-host probability $P(U|x) \approx 0.14$. The second-ranked candidate ($P(O|x) \approx 0.09$) has $z_{\rm phot} = 0.46 \pm 0.04$, somewhat beyond the redshift naively expected from $\mathrm{DM_{excess}} \approx 360$\,pc\,cm$^{-3}$ \citep{Macquart2020}, but still allowed given the large sightline-to-sightline variance in $\mathrm{DM_{cosmic}}$ \citep[e.g.,][]{McQuinn2014}. Neither candidate reaches $P(O|x) \gtrsim 0.99$ threshold for a robust association, so we do not claim a secure host identification.

\subsection{Burst Properties}
\label{sec:properties}

Application of {\tt{fitburst}} burst modeling software \citep{Fonseca2024} provides measurements of burst properties such as the burst's S/N, the dispersion measure (DM), as well as time-of-arrival (TOA) and temporal width of individual sub-bursts and parameters characterizing their spectra. Sub-burst spectra are fit with a well-known model that characterizes the frequency dependent spectral energy distribution as a modified power-law,  
\begin{equation}
    F_{k} = \left(\frac{\nu_k}{\nu_r}\right)^{\gamma + \beta\ln(\nu_k/\nu_r)} \ ,
    \label{eq:fitburst_sed}
\end{equation}
such that $\nu_k$ is the radio frequency for channel $k$, $\nu_r > 0$ MHz is an arbitrary reference frequency and $\gamma, \beta$ are the spectral index and running parameters (see \citet{CHIME/FRB2026a} for details).

We do not see significant evidence for temporal scattering, $\tau$, in any of the bursts from FRB 20210601A. As such, the best-fit profiles displayed in the upper panel of Figure~\ref{fig:waterfalls} are produced from {\tt{fitburst}} models that assume negligible scattering. The absence of significant scattering is further verified with the high time resolution baseband data collected for bursts FRB 20210601C and FRB 20210601F. We determine an upper limit of $\tau\sim0.37$ ms (at 600 MHz) on scattering using the width of the narrowest sub-component (sub-component 3 of FRB 20210601F) plus three times the uncertainty on its width.

\begin{table*}[t]
\begin{center}
\caption{Burst Properties of FRB 20210601A determined from CHIME/FRB intensity data$^a$}
\begin{tabular}{cccccccccc} \hline
TNS Name & S/N$^{b}$ & fluence & DM$^{b}$ & Duration$^{b}$ & Component \#$^{b}$ & TOA$^{c}$ & Width$^{b}$ & $\gamma^{b}$ & $\beta^{b}$ \\
     &     & (Jy ms)  & (pc cm$^{-3}$)  &  (ms)  &     &   (MJD) & (ms)         &          \\ \hline
\hline
\multirow{2}{*}{FRB 20210601A} & \multirow{2}{*}{20.7} & \multirow{2}{*}{18(8)} & \multirow{2}{*}{418.88(5)} & \multirow{2}{*}{21.6(7)} & 1 & 59366.19044981 & 6.7(6) & 23(4) & -104(20) \\
 &  &  &  &  & 2 & 59366.19044996 & 0.77(9) & 5(7) & -56(40) \\
\hline
\multirow{1}{*}{FRB 20210601B} & \multirow{1}{*}{12.8} & \multirow{1}{*}{6(2)} & \multirow{1}{*}{419.2(2)} & \multirow{1}{*}{6.1(5)} & 1 & 59366.19055446 & 2.6(3) & 34(8) & -136(30) \\
\hline
\multirow{1}{*}{FRB 20210601C$^\dagger$} & \multirow{1}{*}{55.2} & \multirow{1}{*}{54(20)} & \multirow{1}{*}{419.6(1)} & \multirow{1}{*}{9.7(2)} & 1 & 59366.19081603 & 4.1(1) & 126(6) & -150(7) \\
\hline
\multirow{1}{*}{FRB 20210601D} & \multirow{1}{*}{18.7} & \multirow{1}{*}{6(3)} & \multirow{1}{*}{419.09(3)} & \multirow{1}{*}{2.04(9)} & 1 & 59366.19101387 & 0.87(6) & 25(4) & -63(10) \\
\hline
\multirow{1}{*}{FRB 20210601E} & \multirow{1}{*}{17.3} & \multirow{1}{*}{7(2)} & \multirow{1}{*}{418.6(1)} & \multirow{1}{*}{6.4(4)} & 1 & 59366.19134189 & 2.7(2) & 49(8) & -103(20) \\
\hline
\multirow{3}{*}{FRB 20210601F$^\dagger$} & \multirow{3}{*}{35.0} & \multirow{3}{*}{24(10)} & \multirow{3}{*}{419.01(4)} & \multirow{3}{*}{13.7(1)} & 1 & 59366.19141217 & 1.31(9) & 103(20) & -94(10) \\
 &  &  &  &  & 2 & 59366.19141225 & 0.49(3) & 207(30) & -178(20) \\
 &  &  &  &  & 3 & 59366.19141230 & 0.82(5) & 73(10) & -69(10) \\
\hline
\multirow{1}{*}{FRB 20210601G} & \multirow{1}{*}{35.1} & \multirow{1}{*}{5(2)} & \multirow{1}{*}{418.4(1)} & \multirow{1}{*}{4.1(3)} & 1 & 59366.19149668 & 1.7(2) & 97(9) & -97(9) \\
\hline
\multirow{2}{*}{FRB 20210601[H,I]$^{d}$} & \multirow{2}{*}{71.6} & \multirow{2}{*}{14(5)} & \multirow{2}{*}{418.0(2)} & \multirow{2}{*}{96.4(3)} & 1 & 59366.19162180 & 5.9(1) & 124(6) & -113(5) \\
 &  &  &  &  & 2 & 59366.19162280 & 2.5(2) & 171(30) & -167(30) \\
\hline
\multirow{1}{*}{FRB 20210601J} & \multirow{1}{*}{16.8} & \multirow{1}{*}{2.1(8)} & \multirow{1}{*}{419.0(2)} & \multirow{1}{*}{12.7(8)} & 1 & 59366.19168271 & 5.4(5) & 28(5) & -57(9) \\
\hline
\end{tabular}
\label{ta:summary}
\end{center}
$^a$ Values in parentheses indicate $1\sigma$ uncertainties on the last quoted digit(s).\\
$^b$ {\tt fitburst} determined parameters.\\
$^c$ Topocentric arrival times referenced at 400.195 MHz.\\
$^d$ 20210601H \& FRB 20210601I are duplicates of the same burst (see Section~\ref{sec:obs}).\\
$\dagger$ Bursts with corresponding baseband data (see Table~\ref{ta:summary_baseband}).
\end{table*}

Properties of individual bursts from FRB~20210601A are summarized in Table~\ref{ta:summary} and plotted versus time in Figure~\ref{fig:prop_vs_time}. While several properties show apparent temporal variability, these measurements carry systematic errors not captured by the quoted formal fitting errors. With respect to DM variations, the narrowband nature of the bursts limits the frequency lever-arm available for DM optimization, and any intrinsic frequency-dependent burst structure introduces further degeneracy between dispersion and spectral morphology. While the two bursts with successful baseband capture (FRB~20210601C and FRB~20210601F) yield DM measurements consistent with the intensity-derived values, the apparent $\sim 0.6$~pc~cm$^{-3}$ DM variation is likely nonphysical, instead arising from degeneracy with frequency-dependent burst structure. The spectral properties $\gamma$, $\beta$ are similarly difficult to interpret, as they are heavily influenced by the chromatic sensitivity of the side-lobe beam response rather than reflecting the intrinsic burst spectrum. This effect is clearly visible in Figure~\ref{fig:beam_map}, where the observed emitting band of each burst closely tracks the frequency-dependent beam sensitivity.

To circumvent these limitations, we construct composite waterfalls for the six bursts for which adjacent beam data are available, selecting at each frequency channel the intensity data from whichever synthesized beam 
provides the highest predicted sensitivity at that frequency and repeating this procedure across the entire $400-800$ MHz bandpass. The resulting composite waterfalls are shown in Figure~\ref{fig:composite_waterfall}. The bursts are clearly broader band than inferred from single-beam data alone, consistent with the spectral properties reported in Table~\ref{ta:summary} being dominated by the beam response rather than the intrinsic spectrum. The enhanced frequency coverage also provides a longer lever-arm for DM optimization. Inspection of the composite waterfalls, all dedispersed to the same weighted-average DM of $\mathrm{DM_{avg}} = 419.01\pm 0.02\;\rm{pc~cm^{-3}}$, shows only weak evidence for DM variation across the burst sequence.

For two bursts for which baseband data were recorded (FRB~20210601C and FRB~20210601F), the chromatic side-lobe response can be more fully circumvented by re-phasing the array to the position of the source (see Section~\ref{sec:localization}), recovering the true sky signal independently of the digitized beam response. The resulting high time-resolution waterfalls are shown in Figure~\ref{fig:baseband_waterfall}. Both bursts reveal significant emission in the lower half of the CHIME band ($400-600$ MHz) that is absent in the single-beam intensity data but consistent with the broader 
spectral coverage recovered from the composite waterfalls, confirming that the intrinsic burst spectra are substantially broader than the single-beam data alone would suggest. The results of applying {\tt fitburst} to baseband data of FRB~20210601C and FRB~20210601F are summarized in Table~\ref{ta:summary_baseband} and displayed in Figure~\ref{fig:prop_vs_time} (red points). Burst properties are generally consistent with those inferred from intensity data (e.g., DM value, insignificant scattering), with some modest but significant differences including an additional subburst component identified for FRB~20210601F and changes in the value of the $\gamma,\beta$ spectral parameters that are consistent with the greater band occupancy of the baseband data.\footnote{Spectral properties are not reported for FRB~20210601C since the burst was fit in ``scintillation mode", which does not fit the burst spectrum with Equation~\ref{eq:fitburst_sed} but instead leaves the per-frequency channel amplitude as a free parameter.}

Baseband data collected for FRB 20210601C and FRB 20210601F also enable the polarized properties of the radio signal to be probed through application of the CHIME/FRB polarization pipeline \citep{Mckinven2021}. We find that both bursts are significantly linearly polarized ($\gtrsim 50\%$). Meanwhile, circular polarization is generally low ($\lesssim 20\%$) and consistent with the level expected from residual instrumental polarization from frequency-dependent phase offsets between CHIME's linearly (X,Y) polarized feeds \citep[see][for details]{Mckinven2021, Pandhi2024}. We detect Faraday rotation measures of $-110.0(2)\; \rm{rad\, m^{-2}}$ and $-111.4(4)\; \rm{rad\, m^{-2}}$ for FRB 20210601C and FRB 20210601F, respectively. Using the best-determined position of the source (see Section~\ref{sec:localization}) and the Faraday rotation sky map of \citet{Hutschenreuter2022}, we estimate the Galactic RM contribution to be $\rm{RM_{MW}}=+50(8)\; \rm{rad\, m^{-2}}$, implying an excess of $\mathrm{RM_{excess}} = \mathrm{RM_{obs}} - \mathrm{RM_{MW}} \sim -150 \text{ to } -170~\mathrm{rad\,m^{-2}}$ in the observer's frame, which expands to $\sim -150 \text{ to } -340~\mathrm{rad\,m^{-2}}$ in the source frame due to cosmological dilation $(1+z)^2$.\footnote{In the absence of a robust host association and redshift measurement, we estimate the maximum redshift of the source and $(1+z)^2$ scaling factor by assuming a simple $\langle \rm{DM_{cosmic}}\rangle\sim 855 \, \textit{z}\; \rm{pc\, cm^{-3}}$ Macquart scaling relation \citep{Macquart2020}, and assuming $\rm{DM_{cosmic}}=\rm{DM_{excess}}\sim 360\; \rm{pc\, cm^{-3}}$, implying a maximum redshift of $\textit{z}_{\rm{max}}\sim 0.4$ for the source.} 

The slight decrease in the observed RM from FRB~20210601C to FRB~20210601F is of marginal significance given the presence of residual instrumental polarization. In particular, non-linear frequency-dependent phase offsets between the X and Y polarization feeds introduce an additional systematic uncertainty which we have estimated elsewhere to be $\sigma_{\rm sys} \sim 0.85~\rm{rad~m^{-2}}$ \citep{Mckinven2023}. Similarly, the narrowband nature of the bursts and the presence of intrinsic frequency-dependent spectral structure introduce systematic uncertainties in the DM measurements that are not captured by the formal fitting errors. As such, the apparent RM and DM variations are unlikely to be significant. 


\subsection{Repetition \& Periodicity Search}
\label{sec:repitition}


The temporal behavior of FRB~20210601A is strikingly clustered: nine bursts detected within a single 106.5 s interval are bracketed by extended intervals of apparent quiescence across multi-year baselines, with no bursts detected in any preceding or subsequent transits. Although the global activity of FRB~20210601A strongly suggests non-stationary behavior, the local 106.5 s active interval shows repetition behavior consistent with a Poisson process. We tested this by comparing the observed waiting-time distribution to 5000 Monte-Carlo realizations of a homogeneous Poisson process with the same mean rate of $\langle \lambda\rangle$ = 0.075 bursts s$^{-1}$. This rate is estimated as $(N-1)/\Delta T$ using the $N-1=8$ inter-arrival intervals spanning the nine detected bursts, rather than $N/\Delta T$, since $\Delta T \approx 106.5$~s is itself defined by the span between the first and last burst and therefore contains no waiting-time information beyond these eight intervals. The observed distribution is statistically consistent with a Poisson process (Monte-Carlo KS test, $\rm{p-value=0.31}$).

We tested how long the observed silence beyond the 106.5~s active interval can persist before it becomes inconsistent with the Poisson rate inferred from the active window itself. Under a homogeneous Poisson process with rate $\lambda$, the probability of observing zero bursts in an additional interval $\Delta T$ is $p(\rm{\Delta T}) = e^{-\lambda \rm{\Delta T}}$. Using the rate $\lambda = \langle\lambda\rangle = 0.075$ bursts s$^{-1}$ estimated from the active window, this probability falls below $p = 0.05$ at $\rm{\Delta T^{*}} \approx 39.9$~s, implying that the observed non-detection beyond $\rm{T_{active}}=106.5$~s is still statistically consistent with the inferred Poisson rate for window extensions up to $\rm{T^{*} = T_{active} + \Delta T^{*}} \approx 146.4$~s.

While we can confidently claim that the source's repetition properties changed after $\rm{T^{*}}$ due to the non-detection of the source in the subsequent main transit, it remains more ambiguous whether the non-detections preceding $\rm{T_{active}}$ reflect an actual change in repetition or whether the active interval was truncated by the significant sensitivity loss at larger angular offsets from meridian. To test this, we implemented a simple simulation framework that bootstraps the observed fluence distribution and detection S/N of the source to produce mock transits of the source 90 minutes prior to the active window. For each trial transit, we draw from the observed fluence distribution and use the CHIME/FRB beam model to translate this into S/N. Due to the significant drop in sensitivity immediately prior to $\rm{T_{active}}$, we are not able to rule out continued activity at fluences $\lesssim 10$~Jy~ms. However, our simulations show that there is sufficient sensitivity in the minutes prior to $\rm{T_{active}}$ that we would expect additional detections from the source at burst fluences $>10$~Jy~ms, had the source's activity extended beyond the apparent $\rm{T_{active}}$ window. From these simulations we conclude that the 106.5~s active interval is not truncated by sensitivity loss in the side-lobes, but instead captures a discrete interval during which the source's repetition properties were distinct.

Complementary to the above repetition analysis, we assessed periodicity using a Lomb--Scargle analysis \citep{Lomb1976,Scargle1982, VanderPlas2015} calibrated via Monte Carlo realizations of a Poisson burst process with identical event counts and duration. The search was conducted over periods ranging from $1.03$~s to $106.52$~s, with periods below the Nyquist limit of $10.29$~s (set by twice the minimum inter-burst interval of $5.14$~s) not reliably interpretable. The strongest observed peak occurs at a period of $5.66$~s with a normalized Lomb--Scargle power of 0.82; 95\% of Monte Carlo Poisson realizations produce a higher maximum power, confirming that the observed peak is entirely consistent with a Poisson null hypothesis. We therefore find no statistically significant periodic modulation over the 106.5 s active window, though we note that the sensitivity of this search is inherently limited by the small number of events ($n=9$) and the narrow range of periods accessible given the duration of the active window.

\subsection{Counterpart Search}
\label{sec:counterpart}

Following the methods of \citet{Curtin2023, Curtin2024}, we searched for any gamma-ray bursts (GRBs) at the location of FRB~20210601A. We limited our GRB sample to those with reported localization uncertainty regions $<3^{\circ}$, as the probability of a coincidence purely due to chance greatly increases with localization size.  While there were no GRBs within a week of the FRB, there are five GRBs with spatial errors less than $3^{\circ}$ that are spatially co-located. These are: GRB950302A \citep[long GRB;][]{1999ApJS..122..465P}, GRB150828A \citep[long GRB; ][]{2020ApJ...893...46V}, GRB151227A \citep[long GRB; ][]{2015GCN.18726....1V, 2020ApJ...893...46V}, GRB161106A \citep[long GRB; ][]{2016GCN.20140....1H, 2020ApJ...893...46V}, and GRB251102A \citep[long GRB; ][]{2025GCN.42531....1F}. However, the event positional uncertainties for these GRBs are close to our threshold (all between $2^{\circ}-3^{\circ}$), and we find that there is a 10\% chance of having five spatial coincidences with these uncertainties. Hence, we do not find any significant GRB coincidences for this FRB. We similarly searched for any X-ray transients found using the Einstein probe \citep{Yuan2018}, and did not find any spatial (within a combined FRB and x-ray transient $5\sigma$ region) coincidences.

At the time of the detected bursts, LIGO was between its O3 and O4 runs, precluding any search for a gravitational wave counterpart. A search for optical transients following the methodology of \citet{Dong2025} likewise yielded no positional coincidences within the $5\sigma$ localization region. Finally, we searched the Very Large Array Sky Survey \citep[VLASS;][]{Lacy2020} catalog for any persistent radio source (PRS) emission associated with FRB~20210601A, finding none and resulting in a $5\sigma$ upper limit of $\lesssim 417.5~\mu$Jy on the $2-4$~GHz flux density of any putative PRS at the source location. For comparison, the brightest known FRB-associated PRSs (FRB~20121102A and FRB~20190520B) have flux densities of $\sim$180-200~$\mu$Jy at GHz frequencies \citep{Chatterjee2017, Niu2022}, meaning our upper limit is a factor of $\sim$2 above the brightest known examples from what remains a small sample of confirmed associations. Deeper radio continuum observations would therefore be necessary to meaningfully constrain PRS emission at the level of known sources.

\section{Discussion}
\label{sec:discussion}

\subsection{Comparison of FRB 20210601A to other CHIME/FRB Repeaters}
\label{discussion:clustering}

\begin{figure*}
	\centering
\includegraphics[width=0.99\textwidth]{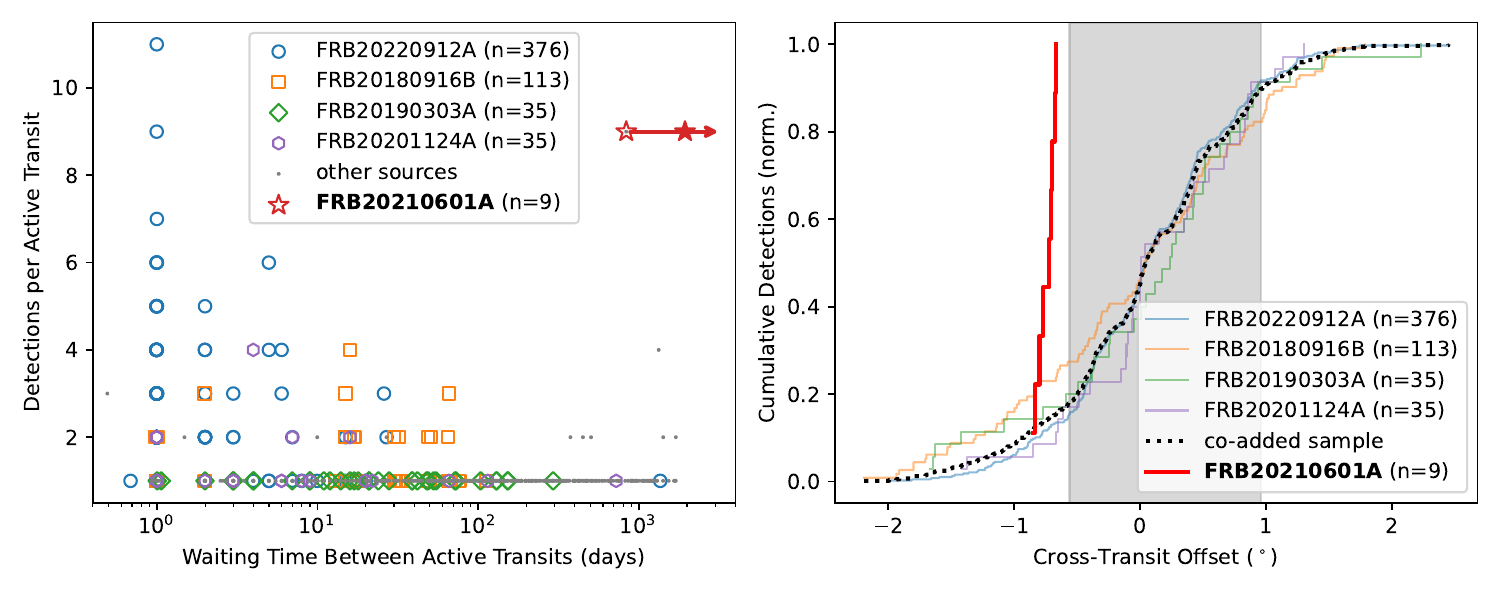}
\caption{Left: Detections per active transit versus waiting time between active transits for CHIME/FRB repeaters. The five most prolific sources are shown with distinct symbols; the remaining CHIME/FRB repeater sample is shown in gray. FRB 20210601A is an extreme outlier: 836 days had elapsed by 2023 September 15 (Catalog 2 end date; open red star) since its sole active transit, increasing to 1674 days by 2025 December 31 (filled red star). The probability of randomly drawing a source from the CHIME/FRB repeater sample with at least as many intra-transit detections and at least as long waiting time between active transits as FRB~20210601A is conservatively estimated to be $\ll 1.6\times10^{-4}$. Right: Cumulative distribution of angular displacement from the CHIME meridian for FRB 20210601A detections (red) compared to a subsample of the most prolific CHIME/FRB repeaters and their co-added distribution (black) of source location at detection. The gray shaded region indicates the range of East-West telescope coordinate, $x$, within which the CHIME/FRB beam sensitivity exceeds 50\% of the peak sensitivity at 600 MHz. An Anderson--Darling (AD) test shows that the angular displacement distribution of FRB 20210601A's detections differs significantly from the co-added distribution of the prolific repeater sample ($p \ll 1\times10^{-3}$).
Repeater properties are drawn from the CHIME/FRB sample presented by \citet{Cook2026}.}
\label{fig:clustering}
\end{figure*}

The nine detections from FRB 20210601A make it among the densest single-transit burst clusters of any repeater observed by CHIME/FRB. While other CHIME/FRB repeaters \cite[e.g., FRB 20220912A \& FRB 20240316A;][]{Mckinven2022, Abbott2024} have shown comparable or higher intra-transit peak activity, they typically are active over longer intervals that span multiple transits. FRB 20210601A distinguishes itself from other sources through its combination of very brief peak activity and subsequent complete apparent quiescence. This is reflected by the novel position that FRB 20210601A occupies relative to other CHIME/FRB repeaters when plotting the number of detections per active transit against waiting time between active transits. This observational parameter space is represented in the left panel of Figure~\ref{fig:clustering}, where FRB~20210601A is indicated by a single data point corresponding to the source's only active transit, with a waiting time lower limit of 836 days that corresponds to the waiting time from 1 June 2021 to the end of the Catalog 2 time frame (15 September 2023). A systematic search of our event database (see Section~\ref{sec:obs}) through December 2025 allows us to further extend the lower limit on the waiting time shown in Figure~\ref{fig:clustering} to 1674 days.

\begin{figure*}
	\centering
\includegraphics[width=0.85\textwidth]{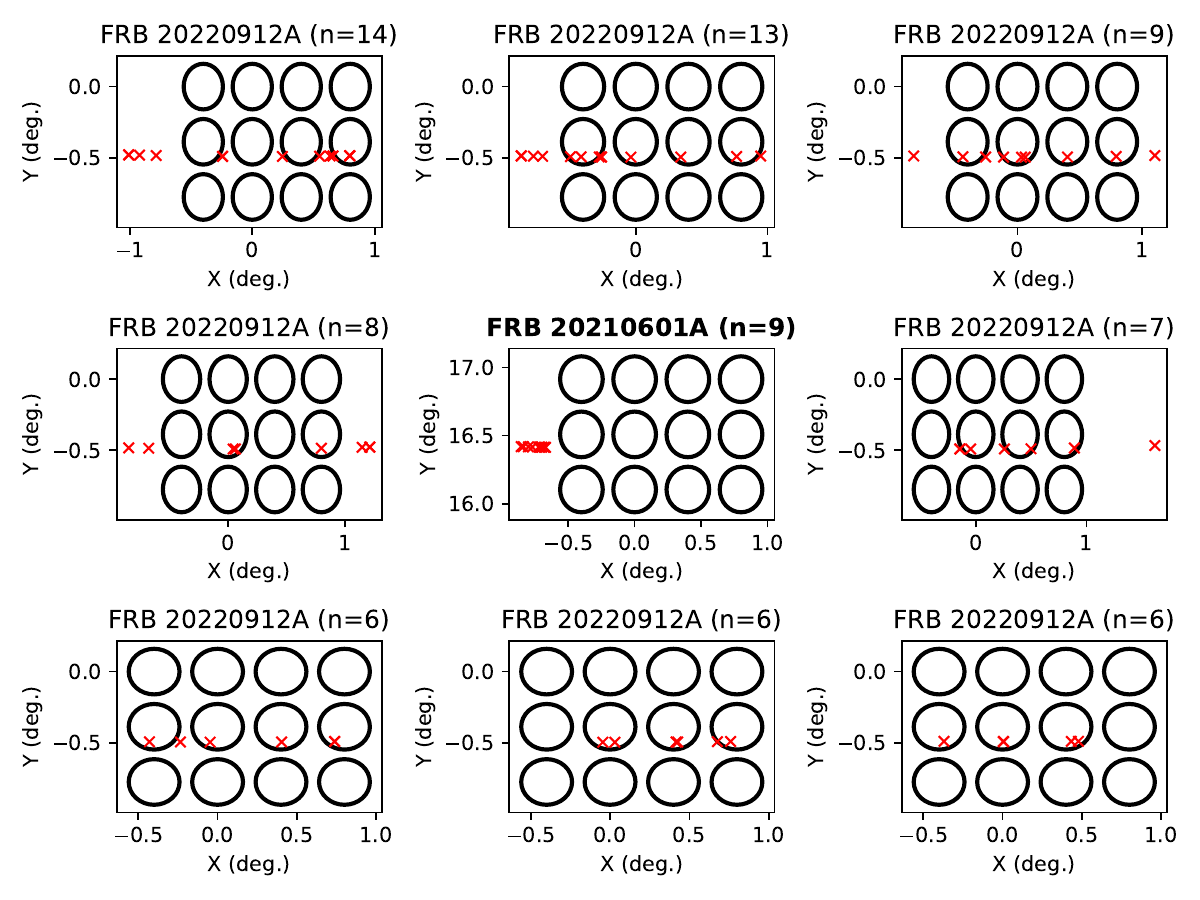}
\caption{Detections as a function of position within the CHIME/FRB FFT beams for FRB 20210601A (centre panel), compared to the most active transits of the prolific repeater FRB 20220912A \citep{Cook2026}. Positions are shown in telescope coordinates (X, Y), where X and Y 
represent the East--West and North--South zenith angles, respectively, with the repeater name and number of detections per transit ($n$) indicated in each panel. Black circles indicate the full-width-half-maximum sensitivity of the FRB search beams at 600~MHz near the source's transit position. Unlike active transits from other CHIME/FRB repeaters, FRB 20210601A detections span a fraction of the source's transit, with all detections occurring off of the main sensitivity peak.}
\label{fig:transit}
\end{figure*}

Unlike typical transits of other CHIME/FRB repeating sources, all nine bursts from FRB~20210601A were detected in a side-lobe of the CHIME/FRB digitized search beams, where the sensitivity is lower and strongly chromatic. The right panel of Figure~\ref{fig:clustering} shows the cumulative distribution of angular offset from meridian for FRB~20210601A's detections alongside those from a sample of prolific CHIME/FRB repeaters. Unlike the prolific repeater subsample, where $\sim75\%$ of detections occur within the main sensitivity lobe of CHIME's FFT beams (gray shaded region), all detections of FRB~20210601A occur outside the main transit. The cross-transit offsets plotted here are sine-projected telescope X (E--W zenith angle) coordinates, which remove the declination dependence of the primary beam width such that the half-power full width is approximately constant across the CHIME sky, enabling a uniform comparison of burst clustering relative to transit regardless of source declination. While the aggregate cumulative distribution in the right panel of Figure~\ref{fig:clustering} demonstrates a strong preference for main-lobe detections across the prolific repeater sample, it remains possible that this trend masks substantial transit-to-transit variability in the location and clustering of individual detections. To test this possibility, we inspected individual transits of hyperactive repeaters and found that, while side-lobe detections do occur, they remain a minority within any given transit, and that active intervals typically span the full duration of the transit. This is strongly contrasted by the tightly clustered, sidelobe-confined pattern observed from FRB~20210601A. These differences are highlighted in Figure~\ref{fig:transit}, which compares the detection pattern of FRB~20210601A's transit to the most active transits of the hyperactive repeating source FRB~20220912A. 

Indeed, FRB~20220912A is a useful source for interpreting FRB~20210601A's anomalous activity, not only serving as a benchmark for the detection pattern of a typical active transit, but also highlighting the significant impact of a source's location relative to CHIME/FRB's static FFT search beams. A dedicated CHIME/Pulsar monitoring campaign of FRB~20220912A, using digitally-formed tracking beams that eliminate the frequency-dependent sensitivity response of CHIME/FRB's static beam grid, recovered 828 bursts over observations conducted in parallel with CHIME/FRB, of which 576 ($\sim70\%$) were missed entirely by CHIME/FRB's static beams during the same transits \citep{Abbott2026}. This demonstrates that static-beam sensitivity nulls, of the kind that likely suppressed detections during FRB~20210601A's sidelobe-confined active window, can substantially undercount a source's true burst rate, further reinforcing the likelihood that FRB~20210601A was more active than the nine detections reported here.

Given that all nine bursts from FRB 20210601A were detected in a side-lobe, we therefore expect that a substantially larger number of bursts would have been detected had the 106.5 s activity window occurred slightly later, when the source was transiting the main-lobe. To estimate the number of bursts that would have been detected in the main-lobe, we average the CHIME/FRB beam response maps over frequency ($400-800$ MHz) and over the four E--W synthesized beams through which the source transits. This yields an effective side-lobe response of $\approx 0.25$ during the activity window, normalized to the main-lobe peak response of $\sim 1.0$. Thus, translating the active window from the side-lobe to the main-lobe implies an average sensitivity gain of $G \approx 1/0.25 = 4$. This effectively lowers the fluence threshold by a factor of $\sim 4$. Using the observed cumulative fluence distribution $N(>F) \propto F^{\rm{index}}$ with $\rm{index} = -0.6(0.2)$ (see Figure~\ref{fig:fluence_dist}), allows us to scale our detections by a factor $\approx 4^{0.6} \approx 2.3$, implying an average of $\sim 21$ bursts would have been detected if the 106.5 s active-window occurred in the main-lobe. This is a conservative lower limit, since the true index is likely significantly steeper than what is reported here due to the selection bias against detection of lower fluence bursts, and because the active interval was likely truncated to an unknown degree prior to the source entering the sensitive region of CHIME's side-lobe. While the bootstrap analysis in Section~\ref{sec:repitition} rules out truncation of the bright-burst population within the sensitivity range tested there, the active interval may still have been truncated to an unknown degree by fainter, sub-threshold bursts prior to the source entering our field of view. If we instead assume a fiducial power-law index of $-1.5$ for the intrinsic energy distribution \citep{abb+23,Curtin2025,Cook2026}, then we would expect $\approx 4^{1.5} \times 9 \approx 70$ detected bursts.\footnote{The above scaling does not take into account the limited frequency coverage of sensitivity in the side-lobe, which acts to further suppress event detection in the side-lobe due to non-uniform sensitivity coverage of the CHIME band.} 

The left panel of Figure~\ref{fig:clustering} reveals a clear anti-correlation between intra-transit burst detections and waiting time between active transits, reflecting the typical activity timescales of repeating sources that span multiple days and are therefore detected across many transits. To estimate the probability of FRB~20210601A occupying its observed location in this parameter space by chance, we conservatively assume independence between these two quantities. Under this assumption, the probability is given by the product of the survival fractions in waiting time between active transits and intra-transit detections relative to the observed CHIME/FRB sample, yielding $p \lesssim 1.6\times10^{-4}$. This should be interpreted as an upper limit on the chance coincidence probability: the observed anti-correlation between intra-transit detections and inter-transit waiting time further suppresses the joint probability of any source occupying the upper-right region of this parameter space, meaning the true probability is likely lower. Moreover, as shown above, the observed burst count from FRB~20210601A could be an underestimate, as its active phase occurred largely outside the primary beam.


Apart from its anomalous repetition, FRB 20210601A exhibits features that are broadly representative of the average repeater behavior across a number of properties. For example, the spectro-morphological properties of FRB 20210601A's bursts are canonically repeater-like \citep[e.g.,][]{Curtin2025}, displaying wide burst envelopes that generally span multiple sub-components and band-limited spectra with complex time--frequency structure, including apparent sub-burst drifting in several events (e.g., FRB~20210601G, FRB~20210601J in Figure~\ref{fig:composite_waterfall}). Burst durations, although generally large, are not anomalous relative to other CHIME/FRB repeaters. Meanwhile, the source's excess dispersion measure ($\rm{DM_{excess}} \sim 360\;\rm{pc\,cm^{-3}}$), excess rotation measure ($|\rm{RM_{excess}}| \sim 150-170\;\rm{rad\,m^{-2}}$), and scattering upper limit $\tau \lesssim 0.37$ ms (at 600 MHz; Section~\ref{sec:results}) all fall close to the average values of the CHIME/FRB repeater sample \citep[e.g.,][]{Curtin2025, Ng2025}. This suggests that FRB~20210601A occupies an environment that is similar to those of other CHIME/FRB repeating sources.

\subsection{Peak Activity and Implications for Energy Budget}

\begin{figure}
\centering
\includegraphics[width=1.0\columnwidth]{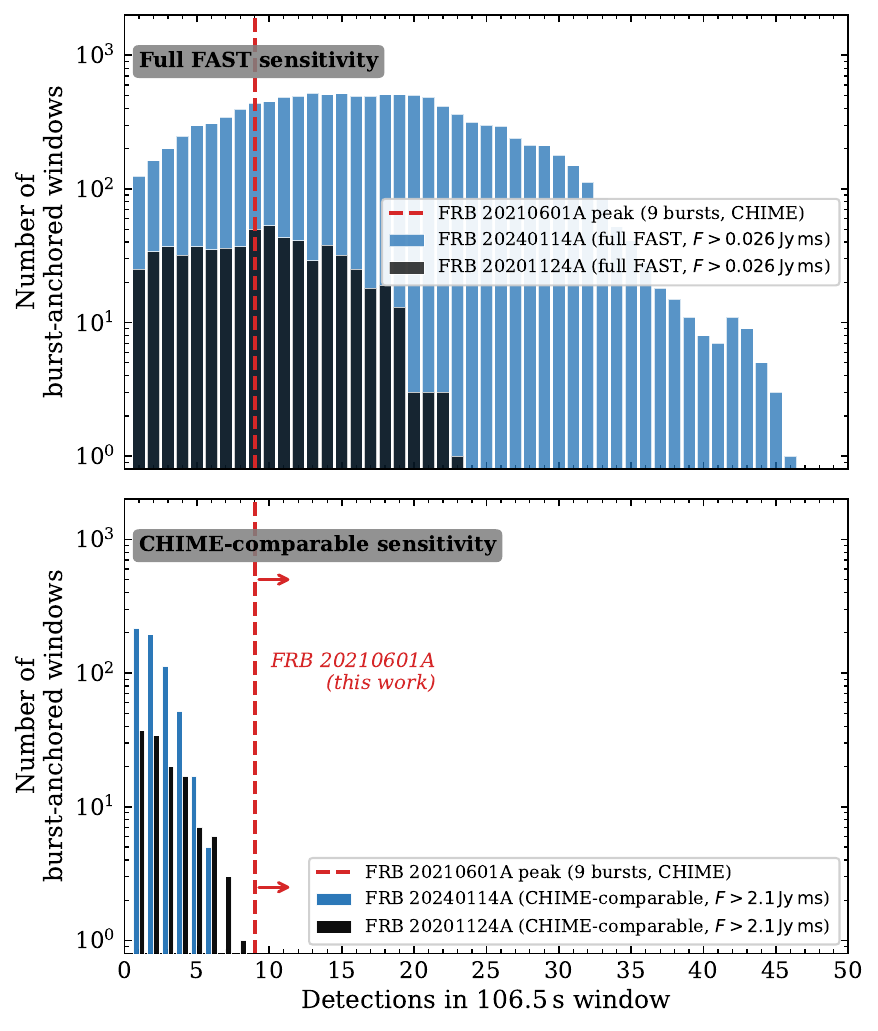}
\caption{Distribution of burst counts per 106.5~s window for FRB~20240114A \citep{Zhang2025} and FRB~20201124A \citep{Zhou2022}, compared to the nine bursts detected from FRB~20210601A (red dashed line). Each entry represents the number of bursts detected in a 106.5~s window anchored at each burst arrival time. \textit{Top panel:} Full FAST sensitivity ($F \gtrsim 0.026$~Jy~ms), showing peak counts of 46 and 23 bursts per window for FRB~20240114A and FRB~20201124A, respectively. \textit{Bottom panel:} Distributions filtered to bursts above the minimum fluence of FRB~20210601A ($F \gtrsim 2.1$~Jy~ms), reducing the peak counts to 6 and 8 bursts, respectively — both below the nine bursts detected from FRB~20210601A. No spectral correction has been applied to account for the differing observing bands of FAST and CHIME/FRB.}
\label{fig:burst_window_distribution}
\end{figure}

To contextualize the exceptional clustering observed during this 106.5 s active window, we compared the burst count distribution of FRB~20210601A to that of the two most prolific repeating FRB sources monitored at much higher sensitivity by the Five-hundred-meter Aperture Spherical Telescope (FAST): FRB~20240114A \citep{Zhang2025} and FRB~20201124A \citep{Zhou2022}. From the full FAST burst catalog of FRB~20240114A (11,553 bursts above $F > 0.026$~Jy~ms, with a peak storm on 2024 March 12) and FAST observations of FRB~20201124A covering a highly active interval \citep{Zhou2022}, we compute the distribution of burst counts within all possible 106.5~s windows anchored at each burst arrival time (Figure~\ref{fig:burst_window_distribution}). At full FAST sensitivity \citep[$F\sim 0.026\; \rm{Jy\, ms}$;][]{Zhang2025}\footnote{This corresponds to the $12\sigma$ fluence completeness limit which we take to be roughly equivalent for both FRB~20240114A \& FRB~20201124A.}, both sources show extraordinarily dense clustering, reaching maxima of 46 and 23 bursts in 106.5 s for FRB~20240114A and FRB~20201124A, respectively. However, when restricted to fluences that exceed the faintest burst detected from FRB~20210601A ($F\gtrsim$ 2 Jy ms), comparable to the sensitivity at which our source was detected, these maxima drop to 6 and 8 bursts, respectively. This comparative analysis is summarized in Figure~\ref{fig:burst_window_distribution} and demonstrates that the nine bursts detected from FRB~20210601A within a single 106.5 s window represent the highest instantaneous burst rate on $\sim$minute timescales yet reported from a repeating FRB source at comparable sensitivity. This is itself a conservative estimate, since the active window occurred while the source was in a CHIME/FRB side-lobe; correcting for the reduced sensitivity implies an equivalent main-lobe rate of $\gtrsim 21$ bursts, exceeding the CHIME-comparable peaks of both comparison sources by more than a factor of two. Such rate comparisons do not account for differing observing bands of CHIME ($400-800$ MHz) and FAST ($1.0-1.5$ GHz) nor the differences in amount and cadence of source exposure. Indeed, CHIME's total exposure to FRB~20210601A ($\sim 320$ hr) significantly exceeds that of FAST in the reported observations of FRB~20240114A and FRB~20201124A ($\sim34$ \& $\sim19$ hr, respectively). However, the FAST monitoring programs for these sources were designed to target periods of elevated activity rather than accumulate untargeted survey exposure as in the case of CHIME. It is therefore unclear whether adjusting to CHIME-equivalent untargeted exposures would increase or decrease the detected burst counts of FRB~20240114A and FRB~20201124A at CHIME-comparable sensitivities.

Following the approach of \citet{Zhang2025}, we estimate the isotropic energy released during this episode and compare it to the dipolar magnetic energy reservoir of a typical magnetar, $E_{\rm{mag}}\sim1.7\times10^{47}$ erg, assuming a surface dipole field strength of $B=10^{15}$ G and a neutron star radius of $R=10^6$ cm. Using the DM-derived maximum redshift ($\textit{z}_{\rm{max}}\sim 0.4$) and the measured fluences of the nine detected bursts, we find $E_{\rm{tot}}\sim2.2\times10^{41}$ erg, or $E_{\rm{tot}}/E_{\rm{mag}}\sim10^{-6}$.\footnote{Using $z_{\rm max}$ and the brightest burst (FRB~20210601C; $F=54(20)$~Jy~ms, width $\approx4.1$~ms), we estimate a maximum isotropic peak luminosity of $L_{\rm iso}\sim2.3\times10^{43}$~erg~s$^{-1}$ ($L_{\rm iso}=4\pi d_L^2 (F/W)\Delta\nu/(1+z)$, $\Delta\nu=400$~MHz), subject to the same redshift caveats as $E_{\rm tot}$ above.} Unlike FRB~20240114A, this estimate carries substantial additional uncertainty along several dimensions: 1) the DM-derived redshift is itself a maximum (assuming zero host DM contribution), 2) the true number of bursts emitted could be considerably higher than nine, both because of the side-lobe sensitivity loss discussed above and because the source may have exhibited other active intervals that went undetected due to CHIME's limited daily exposure, and 3) extrapolating below the detection threshold depends sensitively on both the assumed fluence distribution index and the (unknown) low-energy cutoff. Combining plausible ranges for these factors, $E_{\rm{tot}}/E_{\rm{mag}}$ could reasonably be higher by one to two orders of magnitude, but would remain well below the $\sim$2–87\% constraints derived from the cumulative activity of other prolific repeaters.

\subsection{Interpretation and Analogs}

Strong modulation between active and apparently quiescent phases is not uncommon among repeating FRB sources. For example, FRB~20201124A has shown dramatic variability in its repetition rate, with the source going undetected despite 41.42 hours of CHIME/FRB exposure prior to its detection in November 2020, before entering a month-long episode of intense activity in April 2021 \citep{Lanman2022}. Contemporaneous observations of FRB~20201124A at much higher sensitivity with the Five-hundred-meter Aperture Spherical Telescope (FAST) established a complete quenching of burst activity down to a flux limit of 4.3 mJy on a relatively short timescale \citep[72 hours;][]{Xu2022}. Similarly, FRB 20240114A emerged abruptly in early 2024 with exceptionally high burst rates across a broad range of radio frequencies \citep[][and references therein]{Shin2026}, but its activity has since substantially diminished. Other sources, while perhaps not displaying fully quiescent phases, do display burst storming behavior characterized by highly clustered periods of intense activity interspersed by typically much longer intervals of lower activity \citep[e.g.,][]{Gajjar2018, Nimmo2023}. 

The above examples demonstrate that transitions between active and less active or even fully quiescent states are not unusual among repeating FRBs. However, FRB 20210601A appears to represent an extreme manifestation of this behavior among CHIME/FRB repeaters. Its active phase is confined to a remarkably short interval, during which bursts were clustered at a rate rivaled by only the most active known repeaters, yet this episode was preceded and followed by extended periods of non-detection despite hundreds of hours of exposure. 
The clustered burst activity and subsequent long silence observed from FRB 20210601A may represent an extreme tail of clustered repetition displayed from repeating sources, or may suggest the presence of an entirely new sub-population. In this context, it is worth considering theoretical models and Galactic analogs that may explain the anomalous repetition properties of FRB 20210601A.

\vspace{0.5cm}

\noindent \underline{Nulling/Intermittent Pulsars \& RRAT Analogs:}
Nulling pulsars, intermittent pulsars, and rotating radio transients \citep[RRATs;][]{McLaughlin2006} represent a continuum of increasingly extreme sporadic emission behavior in rotationally powered neutron stars, ranging from pulsars that switch off for individual pulse periods \citep[e.g.,][]{Backer1970, Ritchings1976} to sources detectable only via bright individual pulses separated by intervals of minutes to hours. An extreme example of nulling behavior on much longer timescales is the prototype PSR~B1931+24 \citep{Kramer2006}, which switches between radio-loud and radio-quiet states on timescales of $\sim$5--10 days and $\sim$25--35 days, respectively, with the transition occurring in under 10~s. Across all three classes, the unifying property is access to a highly sporadic magnetospheric emission state with effective duty cycles orders of magnitude lower than those of ordinary pulsars. The parallels with FRB~20210601A are qualitative but suggestive: if the source harbors a neutron star with a long intrinsic period and low duty cycle, its single burst storm could represent the extragalactic, high-luminosity analog of an RRAT activation or intermittent pulsar on-state, rendered detectable only when the source enters its bright emission mode coincidentally with a CHIME/FRB transit. Variability in burst luminosity across activation windows would further suppress the probability of detection, reinforcing the observational impression of extreme episodic behavior.

The recently discovered hyper-luminous RRAT PSR~J2109+50 \citep{Dong2026} provides the most direct Galactic analog identified to date. PSR~J2109+50 exhibits two emission modes: a faint persistent mode detectable via single pulses, and a rare bright mode in which it briefly behaves as a near-persistent periodic emitter with flux densities rivaling the brightest known radio pulsars $\sim$kJy. Monitoring over one month by the CHIME All-sky Multiday Pulsar Stacking Search \citep[CHAMPSS;][]{Andrade2025}, accumulating $\sim$1600~s of exposure per transit, yielded a single detection of the bright mode during a $\sim$90~s interval, implying a bright-mode duty cycle of $\lesssim$0.002. The rarity and brevity of these bright-mode episodes directly echo the extreme temporal confinement of FRB~20210601A's activity window, and demonstrate that such sources are easily missed by surveys with limited dwell times. The key distinction is, however, one of luminosity. PSR~J2109+50's bright-mode bursts are many orders of magnitude lower than the maximum isotropic peak luminosity (i.e., $L_{\rm iso}\sim2.3\times10^{43}$~erg~s$^{-1})$ of this source and typical FRB luminosities. Thus, it remains unknown whether this luminosity gap can be bridged by an extremely radio-luminous analog of PSR~J2109+50.

\vspace{0.5cm}
\noindent \underline{Luminous LPT Analog:}

Long period radio transients \citep[LPTs;][]{Hurley-Walker2022,2025ApJ...990L..49D,2025ApJ...988L..29D,Rea2026} extend the phenomenology of sporadic radio emission from compact objects to much longer periodicities ($\sim$7~min to $\sim$9~hr), occupying a region of $P$--$\dot{P}$ space (where $P$ is the spin period and $\dot{P}$ its time derivative) that disfavors spin-down energy as the dominant power source of their radio emission. While LPTs were originally proposed to be slowly rotating, highly magnetized neutron stars, at least one member of the class has since been conclusively identified as a white dwarf--M-dwarf binary system \citep[e.g.,][]{deRuiter2025}, indicating that the LPT population is likely physically heterogeneous. Unlike nulling pulsars and RRATs, which are plausibly rotationally powered, the neutron-star subset of LPTs likely requires an additional energy reservoir that is suspected to originate from the source's magnetic field, placing them closer to the magnetar regime.

Notwithstanding the large luminosity gap between LPTs and FRBs, this class is relevant to FRB~20210601A in a different sense: LPTs demonstrate that neutron stars can sustain coherent radio emission across a far broader range of spin periods and magnetic field strengths than classical pulsar theory predicts. Some LPT sources additionally exhibit faint ``bridging" emission connecting individual sub-pulses, a morphological feature also seen in some FRBs. Motivated by this, and given the anomalously high activity of FRB~20210601A, one interesting possibility is the existence of additional sub-threshold components or bridging emission connecting its detected sub-bursts. To explore this possibility, we attempted to normalize the detection beam's noise floor using intensity data from neighboring FFT beams when available for each event. This approach was ultimately not sensitive enough to constrain faint bridging emission: near the side-lobes, beam response is highly variable, and the noise level in neighboring beams can be substantially higher than the peak flux density of the burst itself, precluding any meaningful comparison.

\vspace{0.5cm}

\noindent \underline{Magnetar scenarios:}

The April 2020 radio burst from SGR~1935+2154 
\citep{CHIME2020SGR, Bochenek2020} demonstrated that magnetars can emit FRB-like coherent radio bursts, establishing a direct physical link between magnetars and the FRB phenomenon. SGR~1935+2154 subsequently emitted additional, weaker (but still highly luminous) radio bursts on multiple occasions in the months and years following the initial detection \citep[e.g.,][]{Good2020,Kirsten2021,Giri2023}. Notably, two bursts were detected within a single rotational period \citep{Kirsten2021}, highlighting that Galactic magnetars are capable of producing tightly clustered radio emission on sub-rotation timescales. Magnetar starquake models posit that sudden stress release in the crust or magnetosphere triggers avalanche-like burst episodes \citep[e.g.,][]{Thompson1995, Wadiasingh2020, Wang2020, Bransgrove2025}. Such models naturally predict short, intense burst clusters separated by extended quiescent intervals, a phenomenology well-characterized through high-energy light curve analyses \citep{CotiZelati2018}. Interestingly, subsequent radio-active episodes of SGR~1935+2154 have been temporally associated with abrupt changes in the magnetar's spin, including a spin-down glitch coincident with its October 2020 radio activity \citep{Younes2023} and a pair of large spin-up glitches bracketing its October 2022 FRB \citep{Hu2024}. These observations suggest that sudden magnetospheric reconfiguration accompanying a glitch may play a role in triggering short-timescale radio burst episodes. However, their direct relevance to coherent radio emission remains uncertain, as does the extrapolation to the much higher luminosities required for extragalactic detection at typical FRB radio luminosities. Alternatively, magnetar--asteroid interaction models posit that sporadic infall of small bodies onto a magnetar from a surrounding debris disk or disrupted asteroid belt can produce clustered radio bursts \citep{Dai2016, Bagchi2017, Dai2020}.

The high-energy activity observed from Galactic magnetars shares the property of clustered activity: short ($\sim 100$~ms) bursts can cluster into intense ``burst forests'' on timescales of minutes, comparable to FRB~20210601A's activity, albeit in a different energy band and through a physically distinct emission mechanism. These bouts of activity are embedded within broader outbursts of elevated persistent flux spanning hours to days, with subsequent quiescent intervals ranging from weeks to several years \citep[e.g.,][]{Kaspi2017}. The best Galactic analog to FRB~20210601A's specific pattern is 1E~2259+586, which emitted over 80 X-ray bursts within four hours following 5.6~yr of stable, quiescent timing behavior \citep{Kaspi2003}. Even so, the timescale of FRB~20210601A's active window ($\sim$2~min) is markedly more compressed than this or any other known Galactic example, making its combination of an extremely short burst-storm duration and multi-year recurrence lower limit an unusually extreme manifestation of episodic activity. This suggests a distinct triggering mechanism or a magnetospheric state that is both exceptionally short-lived and highly localized.

\vspace{0.5cm}

\noindent \underline{Binary interaction scenarios:}
Periodic or episodic activity may result from a neutron star interacting with the wind or magnetosphere of a binary companion near periastron \citep[e.g.,][]{Ioka2020,Lyutikov2020,Wada2021,Barkov2022}. Such models have been invoked to explain the $\sim16$\,day periodic activity of FRB~20180916B \citep{aab+20,Sand2023} as well as claimed long-term modulation in at least one other repeating source \citep[e.g.,][]{Rajwade2020,Cruces2021}. Broadly, binary interaction scenarios fall into two classes. In ``triggered'' models, FRB emission is intrinsically activated only during a restricted orbital phase, for example when the neutron star traverses a dense stellar wind or magnetospheric structure near periastron \citep[e.g.,][]{Ioka2020, Lyutikov2020, Wada2021}. In contrast, ``passive'' or obscuration models posit that the source is continuously active, but bursts are detectable only during a narrow transparency window set by absorption, scattering, or free–free opacity in the 
circumstellar environment \citep{Lyutikov2020, Barkov2022}. In either case, the observable duty cycle is governed by orbital geometry and line-of-sight propagation effects.

If such a scenario applied to FRB~20210601A, its temporally short activity window would require either a very narrow orbital phase range for intrinsic triggering, or a sharply defined transparency window in an otherwise opaque foreground medium. This could naturally produce clustered bursts tied to orbital phase, though FRB~20210601A has exhibited only a single burst storm to date. Whether an orbital configuration can sustain such behavior without producing additional detectable activity is unknown.

A third possibility is a cataclysmic scenario in which the interacting magnetospheres of two inspiralling neutron stars produce a single, short-lived episode of emission \citep[e.g.,][]{Most2022,Most2023,Cooper2023,Parsons2026}. A testable prediction of such models is that both the burst fluence and the burst rate increase systematically as the orbit shrinks, whereas the inter-burst intervals of FRB~20210601A show no evidence of shortening (Section~\ref{sec:repitition}). Moreover, the detectable window is predicted to be brief ($\sim$ms to seconds before merger), in tension with the $\gtrsim$106.5\,s active interval reported here. Given these tensions, we consider an inspiral-driven scenario disfavored,
although the rapidly varying sidelobe response during the active window limits any test based on fluence evolution.



\vspace{0.5cm}

\noindent \underline{Strong Lensing:}

The burst clustering could, in principle, arise from a transient magnification event rather than intrinsic source activation. Plasma lensing by ionized material has been shown to produce dramatic, short-lived enhancements in single-pulse detection rates. For example, ``black widow" pulsars, where eclipsing material near inferior conjunction with a stellar companion produces caustic-like flux enhancements of order $10^2$ on timescales of seconds \citep{Main2018, Bilous2019}. Moreover, analogous structures in the interstellar medium or local to the source could produce similar effects without requiring a binary companion \citep{SerafinNadeau2025, Zhu2023, Fiedler1987}. In such a scenario, relative motion between the source, lens, and observer naturally produces a brief magnification window, rendering an otherwise sub-threshold repeating source temporarily detectable. Assuming a relative velocity of $\sim$100~km~s$^{-1}$, the 106.5~s active 
window corresponds to a crossing scale of $\sim$0.02~$R_\odot$, consistent with the Einstein radius of a $\sim$0.2~$M_\odot$ companion or the Fresnel scale of a compact plasma structure, both at $\sim$1~AU separation.

However, plasma lensing is expected to produce strongly chromatic amplification and, in some cases, enhanced temporal scattering \citep{Cordes2017}. Neither of these propagation effects can be robustly assessed given the chromatic side-lobe beam response and the absence of detected scattering. We therefore consider lensing a physically motivated but currently unconstrained explanation for the observed burst clustering.

\subsection{Poisson Limits on Episodically Active Repeater Population}
\label{discussion:modelling}

\begin{figure}
	\centering
\includegraphics[width=0.5\textwidth]{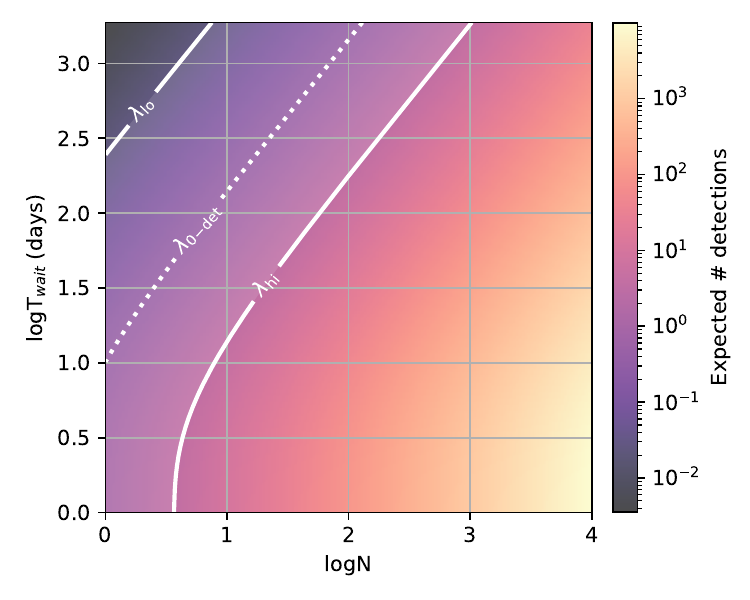}
\caption{Constraints on the episodic repeater population parameters $N$ (total population size) and $T_{\rm{wait}}$ (storm recurrence time), with storm duration fixed to the observed burst clustering window of FRB~20210601A, $T_{\rm{storm}} = 106.5$~s. The background heatmap shows the expected number of detections, $\Lambda(N, T_{\rm{wait}})$, evaluated at fixed $T_{\rm{storm}}$. The 
solid white contours, labeled $\Lambda_{\rm lo}$ and $\Lambda_{\rm hi}$, bracket the 95\% interval of $\Lambda$ evaluated over the one-detection posterior samples (i.e., 
the 2.5th and 97.5th percentiles); since the Poisson likelihood for a single detection peaks at $\Lambda = 1$ and declines on both sides, this scenario excludes both 
overly large and overly small expected detection rates. The dotted contour, labeled $\Lambda_{\rm 0\text{-}det}$, marks the 95th-percentile $\Lambda$ value of the zero-detection posterior, representing a one-sided upper limit. The single detection of FRB~20210601A favors moderate population sizes but is degenerate with $T_{\rm{wait}}$, which remains largely unconstrained by the data, as reflected by the near-vertical extent of both contours across the full sampled range.}
\label{fig:exclusion}
\end{figure}

Motivated by the anomalous repetition properties of FRB~20210601A, we construct a simple toy model for a population of episodically active repeaters displaying 
clustered activity. This framework makes no assumptions about the underlying physical source model and is purely a characterization of repetition statistics and survey 
properties of CHIME/FRB. As such, we do not attempt to constrain the volumetric rate or luminosity function of the population. Rather, we ask how many sources displaying FRB~20210601A-like activity could exist within the sensitivity horizon of CHIME/FRB, and whether their non-detection elsewhere in the survey is consistent with the model.

To constrain the properties of a population of repeating FRBs exhibiting temporally clustered emission, we adopt a Poisson likelihood framework that models the expected number of detections across the sky accounting for the declination-dependent exposure of CHIME/FRB. We adopt a simplified framework in which each source is assumed to 
produce bursts in episodic ``storms'' of duration $T_{\rm{storm}}$, repeating every $T_{\rm{wait}}$ days, over the survey duration $T_\mathrm{survey}$. The model is governed by two free parameters:
\begin{itemize}
    \item $N$: the total number of clustered repeaters on the sky,
    \item $T_{\rm{wait}}$: the average spacing between burst storms (in days).
\end{itemize}
The storm duration $T_{\rm{storm}}$ is fixed to the observed burst clustering window of FRB~20210601A, $T_{\rm{storm}} = 106.5$~s, rather than inferred from the data. We find that $T_{\rm{wait}}$ and $T_{\rm{storm}}$ are nearly degenerate in the likelihood: for fixed $N$, increasing $T_{\rm{wait}}$ can be compensated by decreasing $T_{\rm{storm}}$ to maintain the same expected detection rate, leaving both parameters unconstrained by survey statistics alone when fitted simultaneously. Since 
$T_{\rm{storm}}$ is directly measured from the observed burst sequence of FRB~20210601A, fixing it breaks this degeneracy and reduces the inference to a well-constrained two-parameter problem.

We adopt a physically motivated prior on $T_{\rm{wait}}$, restricting it to the range $[1~\mathrm{day},\, T_{\rm{survey}}]$. The lower bound of one day ensures model validity, as the exposure model $d(\delta)$ is defined as a daily average and becomes 
ill-defined when multiple storms occur within a single dwell window ($T_{\rm{wait}} < 1$~day). The upper bound of $T_{\rm{survey}} \approx 1875$~days reflects the fact that sources with inter-storm intervals exceeding the survey 
duration contribute negligibly to the expected detection rate and are therefore unconstrained by the data.

For each declination $\delta$, we compute the expected observing time per day, $d(\delta)$ (in minutes), using an empirical model of telescope exposure informed from the declination-dependent exposures reported by \citet{CHIME/FRB2026a}. Namely,
\begin{equation}
d(\delta) = \frac{E_0}{\cos(\delta)},
\label{eqn:exposure}
\end{equation} 
where $E_0\sim2.2$ minutes is the baseline-level of per day exposure of low declination sources.

The probability of detecting at least one burst during the survey interval, $T_\mathrm{survey}$ (e.g., the 1875 day interval of CHIME/FRB Catalog 2), from a single source at declination $\delta$ is given by
\begin{equation}
    p_\mathrm{det}(\delta) = 1 - \left(1 - \frac{d(\delta) + T_{\rm{storm}}}{1440~\mathrm{min}} \right)^{n_\mathrm{storms}},
\end{equation}
where $n_\mathrm{storms} = T_\mathrm{survey} / T_{\rm{wait}}$ is the total number of burst storms per source. The term in parentheses represents the daily probability of overlapping a storm with the telescope dwell time.

The expected number of detections in each declination bin is then
\begin{equation}
    \Lambda(\delta) = N \times p_\mathrm{det}(\delta) \times w(\delta),
\label{eq:lambda}
\end{equation}
where $w(\delta) \propto \cos(\delta)$ is a sky-area weighting factor normalized across all declination bins.

Assuming independent Poisson statistics across the sky, the likelihood of observing $k_i$ detections in each declination bin $i$
\begin{equation}
    \log \mathcal{L} = \sum_i \left[ k_i \log \Lambda_i - \Lambda_i - \log(k_i!) \right],
\end{equation}
where $k_i \in \{0,1\}$ corresponds to the number of observed clustered repeaters in bin $i$. We consider two scenarios:
\begin{enumerate}
    \item \textbf{One-detection case}: A single clustered repeater detection in one 
    bin (i.e., detection of FRB 20210601A at declination $\sim$65$^\circ$) and zero 
    elsewhere. This scenario constrains the parameter space to models that predict a 
    small number of detections.
    \item \textbf{Zero-detection case}: No clustered repeaters are detected in any 
    declination bin, placing stricter bounds on the credible regions of all model 
    parameters, $N$ and $T_{\rm{wait}}$.
\end{enumerate}

For the one-detection scenario, the likelihood is evaluated bin-by-bin across the declination grid. In the bin corresponding to the declination of FRB~20210601A 
($\delta \approx 65^{\circ}$), we adopt a Poisson likelihood for a single detection,
\begin{equation}
    \log \mathcal{L}_{\rm 1-det} = \sum_{i \neq i_{\rm det}} \left(-\Lambda_i\right) 
    + \log\Lambda_{i_{\rm det}} - \Lambda_{i_{\rm det}},
    \label{eq:loglike_1det}
\end{equation}
where $\Lambda_i = N \times p_{\rm det}(\delta_i) \times w(\delta_i)$ is the expected number of detections in bin $i$ (Equation~\ref{eq:lambda}), and $i_{\rm det}$ denotes 
the bin containing FRB~20210601A. The first sum corresponds to $\log P(k{=}0\,|\,\Lambda_i) = -\Lambda_i$ for all non-detection bins, while the 
second term corresponds to $\log P(k{=}1\,|\,\Lambda_i) = \log\Lambda_i - \Lambda_i$ for the detection bin. This formulation naturally drives the posterior toward parameter combinations for which the sky-integrated detection rate is of order unity, $\sum_i \Lambda_i \sim 1$.

We use the \texttt{UltraNest} \citep{Buchner2021} nested sampling algorithm to infer the posterior distribution over $\log_{10}(N)$ and $\log_{10}(T_{\rm{wait}})$, with 
$T_{\rm{storm}}$ fixed to the observed value. This approach enables quantitative constraints on the number and activity pattern of clustered repeaters that are consistent with observed survey statistics.

Rather than summarizing the posterior with a kernel-density-based credible region, we exploit the fact that the deterministic detection rate $\Lambda(N, T_{\rm{wait}})$ is the quantity directly entering the likelihood, and define exclusion contours as level sets of $\Lambda$ evaluated at percentiles of its distribution over the posterior samples. For the one-detection scenario, the Poisson likelihood $\log\mathcal{L}_{\rm 1-det} \propto \log\Lambda - \Lambda$ peaks at $\Lambda = 1$ and declines for both larger and smaller $\Lambda$; we therefore bracket the posterior with two contours, $\Lambda_{\rm lo}$ and $\Lambda_{\rm hi}$, corresponding to the 2.5th and 97.5th percentiles of $\Lambda$ evaluated over the one-detection posterior samples, together enclosing a 95\% interval. For the zero-detection scenario, the likelihood $\log\mathcal{L}_{\rm 0-det} \propto -\Lambda$ is monotonically decreasing in $\Lambda$, so a single one-sided contour at the 95th percentile of $\Lambda$ over the zero-detection posterior samples, $\Lambda_{\rm 0\text{-}det}$, suffices as an 
upper limit.

The resulting constraints are shown in Figure~\ref{fig:exclusion}. With $T_{\rm{storm}}$ fixed to the observed value, the single detection of FRB~20210601A 
favors models with moderate population sizes ($\log N \sim 1$--$2$) over a wide range of storm recurrence times. The diagonal orientation of the $\Lambda_{\rm lo}$ and 
$\Lambda_{\rm hi}$ contours reflects a residual degeneracy between $N$ and $T_{\rm{wait}}$: at fixed $T_{\rm{wait}}$, $N$ is tightly bracketed by the data, but marginalizing over the full prior range of $T_{\rm{wait}}$ leaves the recurrence time itself essentially unconstrained. The zero-detection scenario provides a complementary, more restrictive upper limit on $N$ for a given $T_{\rm{wait}}$, consistent with the absence of similar clustered repeaters elsewhere in the survey.

\subsubsection{Forecasting CHIME/FRB Detections of Episodic Repeater Population}

\begin{figure}
	\centering
\includegraphics[width=0.45\textwidth]{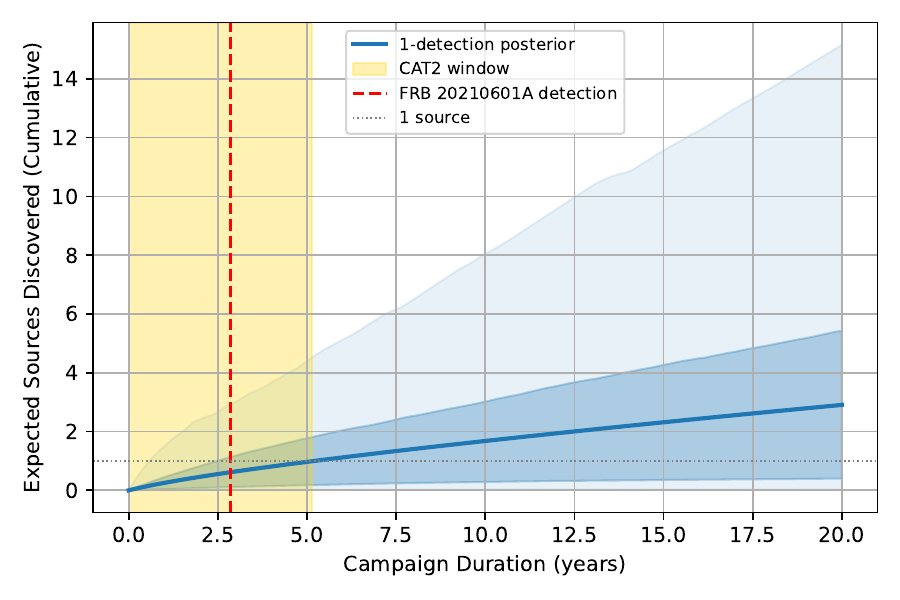}
\caption{Forecasted cumulative number of detected episodic repeaters as a function of CHIME/FRB observing campaign duration. The solid line shows the mean prediction from the 1-detection posterior, with dark and light shaded regions indicating 68\% and 99\% prediction intervals obtained by marginalizing over the posterior distributions of $(N, T_{\rm{wait}})$, with $T_{\rm{storm}}$ fixed to the observed value. The yellow shaded region marks the CHIME/FRB Catalog 2 duration (2018 July 25 to 2023 September 15), with the red vertical line indicating the detection epoch of FRB~20210601A at $\sim2.85$ years.}
\label{fig:detections_vs_time}
\end{figure}

To connect the inferred repetition model parameters to observational outcomes, we sample from the posterior distributions of Figure~\ref{fig:exclusion}. In particular, for each sample from $(N, T_{\rm{wait}})$, with $T_{\rm{storm}}$ fixed to the observed value, we compute the detection probability as a function of campaign duration, taking into account the declination–dependent exposure of CHIME/FRB and sky–averaged solid–angle weighting. Multiplying this detection probability by the total population size yields the expected number of detected repeaters for that parameter draw (heatmap of Figure~\ref{fig:exclusion}). 

By marginalizing over the full posterior ensemble, we obtain a prediction on the distribution of the number of detections as a function of survey campaign duration, which can be directly compared to the actual outcome of the Catalog 2 interval. The resulting posterior-marginalized trend is shown in Figure~\ref{fig:detections_vs_time}. The mean prediction at the end of the Catalog~2 interval ($\sim 5.14$~years) is $\langle N_{\rm det} \rangle \sim 1$, recovering the single detection of FRB~20210601A by construction rather than as an independent validation.

Although the mean prediction agrees with observations, the single detection provides only weak constraints on the model parameters. The inferred total size of the episodic repeater subpopulation detectable by CHIME/FRB is consequently highly uncertain: $\rm{N}=9^{+200}_{-8}$ sources ($68\%$ credible interval). This uncertainty, combined with the largely unconstrained marginal posterior on $T_{\rm{wait}}$ (Section~\ref{discussion:modelling}), is reflected in the wide shaded regions of Figure~\ref{fig:detections_vs_time}.

The model predicts the detection rate of sources entering episodic burst storms. Classifying a source as an episodic repeater requires multiple burst detections within a storm window, which depends on the intra-storm burst rate -- a parameter not constrained by our current model. The forecast therefore represents an optimistic upper bound on the discovery rate of sources like FRB~20210601A. Indeed, it remains possible that some one-off FRBs or less active repeaters discovered by CHIME/FRB actually display intense, temporally confined burst storms similar to FRB~20210601A, but are not identified as such because their active intervals only partially overlap CHIME's exposure window, or because sensitivity thresholds limit detections to only the highest-fluence, but less common, bursts. Future analyses incorporating more flexible repetition models and a refined treatment of CHIME/FRB's declination and frequency-dependent sensitivity may further improve the constraints provided here as would a greater number of source discoveries from this putative sub-population.

\section{Conclusions}
\label{sec:conclusions}

We report the discovery of FRB~20210601A which exhibits repetition properties that interestingly distinguish it from the broader CHIME/FRB repeater population.

\begin{itemize}
    \item \textbf{Extreme peak activity:} The source produced at least nine bursts within a fraction of a single CHIME transit, making it the most densely clustered single-transit burst episode observed by CHIME/FRB. While other repeaters (e.g., FRB~20220912A, FRB~20240316A) have exhibited comparable intra-transit activity levels, such intervals of enhanced activity are distributed over longer timescales spanning multiple transits. Even when compared against FAST observations of hyperactive repeaters such as FRB~20240114A and FRB~20201124A, the 106.5 s active interval of FRB~20210601A represents the highest instantaneous burst rate yet observed from a source at CHIME-comparable fluence thresholds.

    \item \textbf{Tight temporal confinement followed by prolonged inactivity:} No bursts have been detected before or after this storm, despite continued CHIME/FRB monitoring. In contrast, most repeaters display either recurring active windows or sporadic low-level activity across many epochs. This combination of peak activity and extended interval of inactivity is anomalous relative to the CHIME/FRB repeater sample.

    \item \textbf{Population implications:} The existence of such a source can naturally be explained by a population of repeaters characterized by rare, short-duration burst storms separated by long recurrence times. Such a putative population is poorly constrained by most surveys, but is preferentially accessible to CHIME/FRB owing to its high cadence and wide sky coverage.

\end{itemize}

Whether FRB~20210601A represents a statistical outlier or instead the first clear example of a previously under-sampled class of highly episodic repeaters remains uncertain. The analysis presented here demonstrates that sources with brief activity windows (tens of minutes) and long quiescent intervals (weeks-to-months or longer) inhabit a region of parameter space that remains largely unconstrained by existing surveys. 
If a population of infrequently active repeaters exists, it should preferentially be discovered in regions of the sky with the highest combined sensitivity and exposure. The detection of FRB~20210601A near CHIME’s zenith, where the instrument sensitivity is maximal and daily exposure remains substantial, is consistent with this expectation.  Future discoveries from this population should therefore also preferentially cluster in sky regions with similarly optimal combinations of sensitivity and exposure.

The highly clustered emission observed from this source implies a strongly non-Poissonian/non-stationary and episodic emission process. Several physical interpretations and Galactic analogs have been explored in this paper. One possibility is that the bursts arise from short-lived magnetospheric instabilities in a magnetar, an extension of the X-ray/$\gamma$-ray burst storms seen from some Galactic magnetars but at radio wavelengths. However, this analogy remains speculative for the radio band, as dedicated monitoring during high-energy bursting episodes (e.g., SGR J1935+2154)  has not established that individual X-ray bursts are reliably accompanied by coherent radio counterparts \citep[e.g.,][]{Lin2020,Ibrahim2024}. Alternatively, the behavior may resemble an extreme manifestation of nulling or RRAT-like intermittency, in which rapid transitions between radiatively efficient and inefficient magnetospheric states produce temporally localized detectability. Binary interaction scenarios, including both phase-triggered emission and orbital transparency-window models, could also generate short activity windows, though such models often predict some degree of periodic modulation and/or DM or scattering variations that are not clearly evident here. 

At present, no single Galactic analog or theoretical framework provides a compelling one-to-one match to the observed phenomenology. While several candidate analogs can plausibly reproduce the repetition behavior in isolation, whether any of them can simultaneously account for other defining properties of FRBs, notably their extreme radio luminosities, remains highly speculative and discriminating among competing explanations is difficult given a single detection of this kind of source. Regardless of the underlying physical mechanism, FRB~20210601A demonstrates that the repetition properties of FRB sources span a broader range than previously established. Targeted high-sensitivity follow-up of this source, for example with FAST or other large single-dish instruments, offers the opportunity to detect fainter bursts ($\ll$ Jy ms) during quiescent periods and could help rule out the episodic hypothesis explored here. Meanwhile, continued untargeted searches with wide field-of-view surveys provide the exposure necessary to discover additional members of this potentially hidden population of highly episodic repeaters.

\begin{acknowledgments}

We acknowledge that CHIME is located on the traditional, ancestral, and unceded territory of the Syilx/Okanagan people. We are grateful to the staff of the Dominion Radio Astrophysical Observatory, which is operated by the National Research Council of Canada. CHIME operations are funded by a grant from the NSERC Alliance Program and by support from McGill University, University of British Columbia, and University of Toronto. CHIME was funded by a grant from the Canada Foundation for Innovation (CFI) 2012 Leading Edge Fund (Project 31170) and by contributions from the provinces of British Columbia, Québec and Ontario. The CHIME/FRB Project was funded by a grant from the CFI 2015 Innovation Fund (Project 33213) and by contributions from the provinces of British Columbia and Québec, and by the Dunlap Institute for Astronomy and Astrophysics at the University of Toronto. Additional support was provided by the Canadian Institute for Advanced Research (CIFAR), the Trottier Space Institute at McGill University, and the University of British Columbia. The CHIME/FRB baseband recording system is funded in part by a CFI John R. Evans Leaders Fund award to IHS. FRB research at UBC is funded by an NSERC Discovery Grant and by the Canadian Institute for Advanced Research. The AstroFlash research group at McGill University, University of Amsterdam, ASTRON, and JIVE is supported by: a Canada Excellence Research Chair in Transient Astrophysics (CERC-2022-00009); an Advanced Grant from the European Research Council (ERC) under the European Union's Horizon 2020 research and innovation programme (`EuroFlash'; Grant agreement No. 101098079); an NWO-Vici grant (`AstroFlash'; VI.C.192.045); an NSERC Discovery Grant (RGPIN-2025-06681); an ERC Starting Grant (`EnviroFlash'; Grant agreement No. 101223057); and an NWO-Veni grant (VI.Veni.222.295).

A.B.P.~acknowledges support by NASA through the NASA Hubble Fellowship grant \mbox{HST-HF2-51584.001-A} awarded by the Space Telescope Science Institute, which is operated by the Association of Universities for Research in Astronomy, Inc., under NASA contract \mbox{NAS5-26555}. A.B.P.~also acknowledges prior support from a Banting Fellowship, a McGill Space Institute~(MSI) Fellowship, and a Fonds de Recherche du \mbox{Qu\'ebec -- Nature} et Technologies~(FRQNT) Postdoctoral Fellowship.
A.M.C. is a Banting Postdoctoral Research Fellow. 
A.P.C. is a Canadian SKA Scientist and is funded by the Government of Canada / est financé par le gouvernement du Canada.
A.P. is a Trottier Space Institute Postdoctoral Fellow.
C. L. acknowledges support from the Miller Institute for Basic Research at UC Berkeley.
E.F. is supported by the National Science Foundation under grant AST-2407399.
F.A.D is a Canadian SKA Scientist and is funded by the Government of Canada / 
est financé par le gouvernement du Canada.
H.M.T. is a Trottier Space Institute Postdoctoral researcher.
K.N. acknowledges support by NASA through the NASA Hubble Fellowship grant \# HST-HF2-51582.001-A awarded by the Space Telescope Science Institute, which is operated by the Association of Universities for Research in Astronomy, Incorporated, under NASA contract NAS5-26555
K.T.M is supported by a FRQNT Master’s Research Scholarship.
K.W.M. is supported by NSF Grant No. 2510771 and receives lumbar support from the Adam J. Burgasser Chair in Astrophysics. 
M.N. is a Fonds de Recherche du \mbox{Qu\'ebec -- Nature} et Technologies (FRQNT) postdoctoral fellow.
M.W.S is a Fonds de Recherche du \mbox{Qu\'ebec -- Nature} et Technologies (FRQNT) postdoctoral fellow and acknowledges support from the Trottier Space Institute Fellowship program. 
P.S. acknowledges the support of an NSERC Discovery Grant (RGPIN-2024-06266).
V.M.K. holds the Lorne Trottier Chair in Astrophysics \& Cosmology, a Distinguished James McGill Professorship, and receives support from an NSERC Discovery grant (RGPIN 228738-13).

\end{acknowledgments}

\bibliographystyle{aasjournalv7}

\bibliography{frbrefs,extra}

\newpage
\appendix
\renewcommand{\thefigure}{A.\arabic{figure}}
\renewcommand{\thetable}{A.\arabic{table}}
\setcounter{figure}{0} 
\setcounter{table}{0}
\renewcommand{\theHfigure}{A.\arabic{figure}}
\renewcommand{\theHtable}{A.\arabic{table}}

\section{Supplemental Figures \& Tables}
\label{appendix:supplemental}

\begin{figure*}[h]
	\centering
\includegraphics[width=1.0\textwidth]{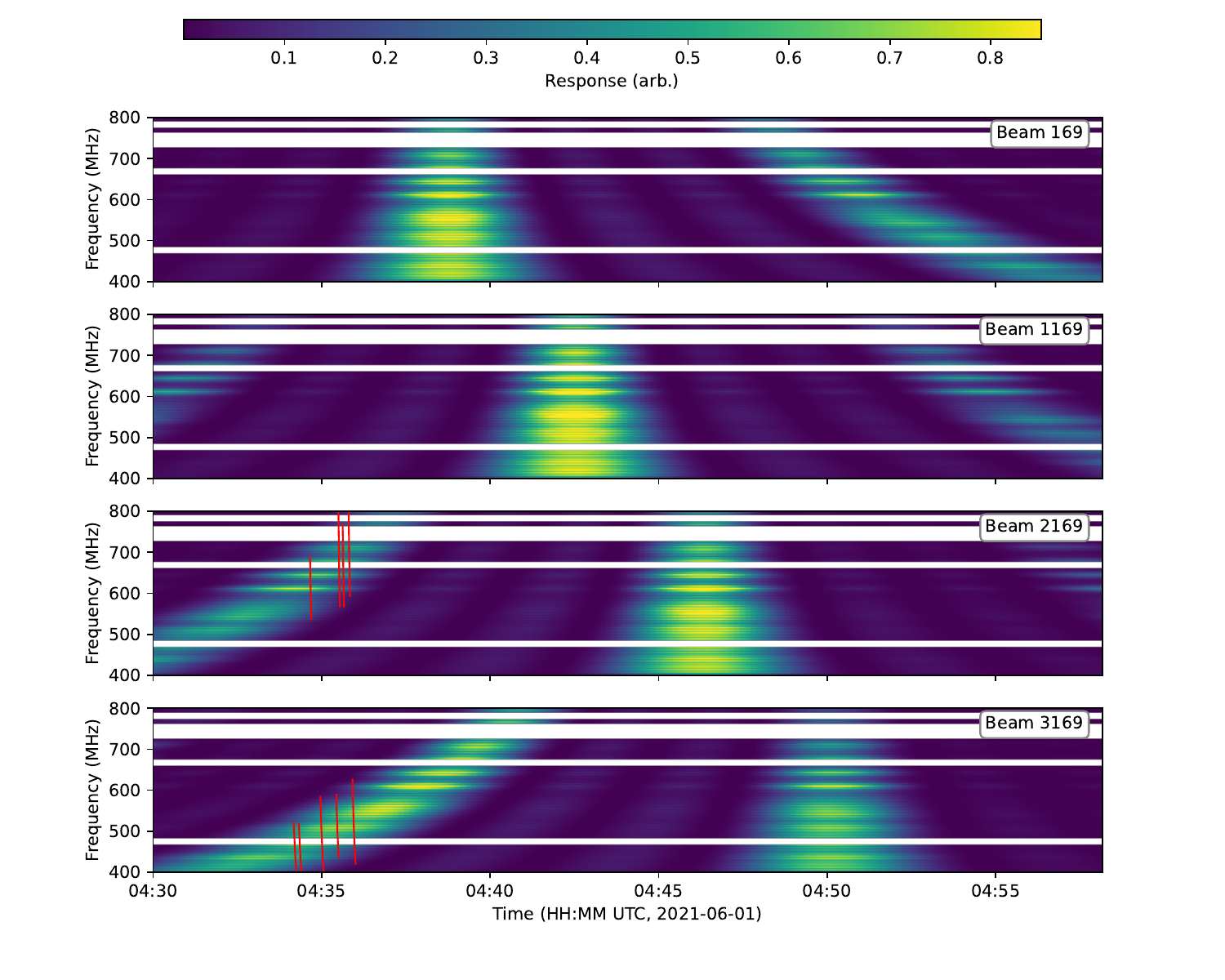}
\caption{Detections of bursts from FRB 20210601A as a function of time (UTC) relative to the CHIME/FRB beam model. The associated color map shows beam sensitivity (arbitrary units) for each of the four beams along the source's transit path. The emitting band of individual bursts is indicated by red lines at their topocentric TOA; these are not perfectly vertical, since they account for the dispersive delay across the band from the burst's non-zero DM. This highlights that the observed spectral properties of each burst are dominated by the chromatic sensitivity of the beam.}
\label{fig:beam_map}
\end{figure*}

\begin{figure*}[h]
	\centering
\includegraphics[width=0.8\textwidth]{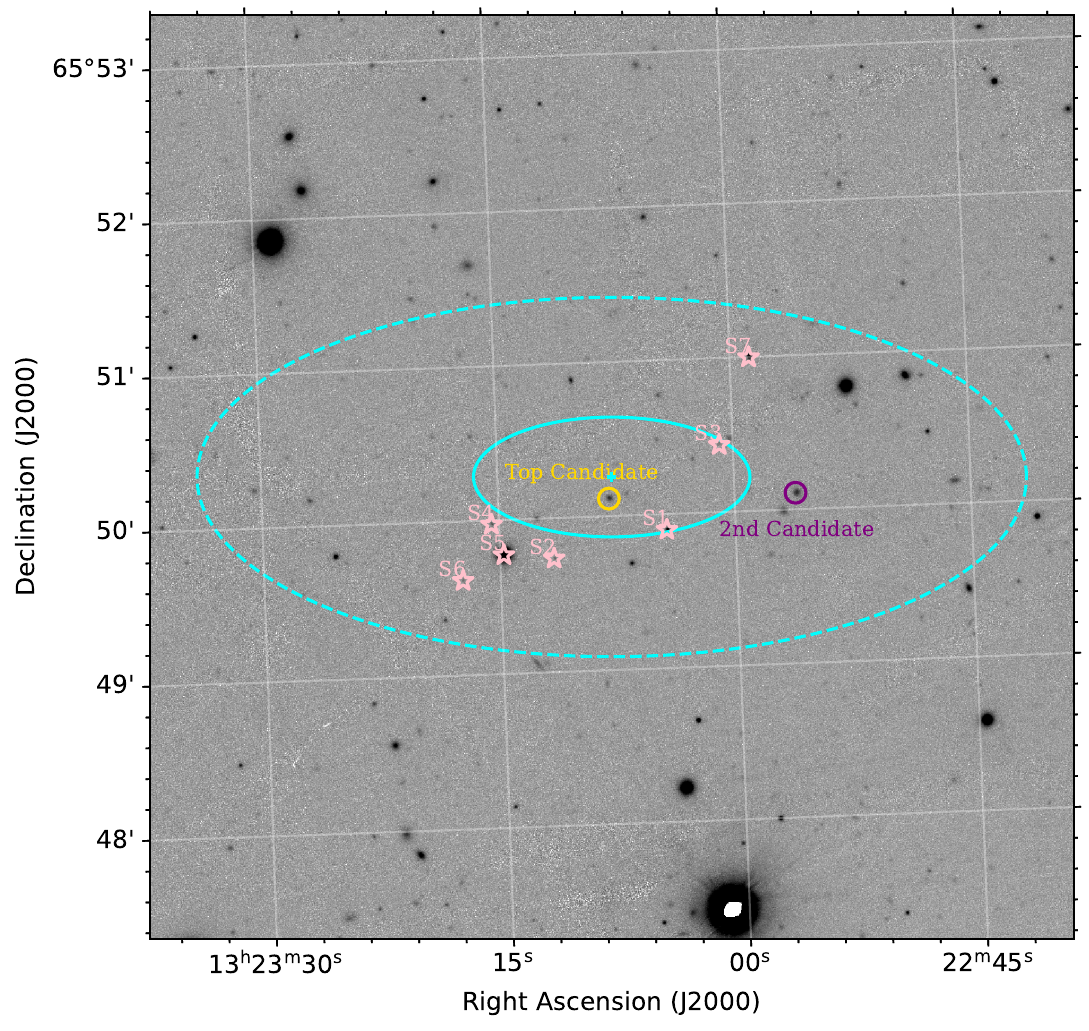}
\caption{Pan-STARRS1 3$\pi$ $i_{\mathrm{P1}}$-band stacked image \citep{chambers2016pan,Waters16} of the approximate localization region of FRB 20210601A, represented by $1\sigma$ (solid) \& $3\sigma$ (dashed) error ellipses. Foreground stars are indicated by pink star markers taken from the Gaia catalog \citep[DR3;][]{Gaia2023}. The image corresponds to the unconvolved ($\texttt{unconv}$) stack, preserving the native point-spread function and providing deep optical imaging of the field with a pixel scale of $\sim$0.25'' per pixel. Host candidates and their PATH posterior probabilities were derived from the DESI Legacy Imaging Surveys DR9 (BASS/MzLS) catalog \citep{Dey2019}; the two highest-ranked candidates are marked, with $P(O|x) \approx 0.14$ (Top Candidate) and $P(O|x) \approx 0.09$ (2nd Candidate).}
\label{fig:localization}
\end{figure*}

\clearpage


\startlongtable
\begin{deluxetable*}{ccccccccccc}
\tablewidth{0pt}
\tablecaption{Burst properties of FRB 20210601A determined from CHIME/FRB baseband subsample$^{a}$
\label{ta:summary_baseband}}
\tablehead{
\colhead{TNS Name} & \colhead{S/N$^{b}$} & \colhead{fluence} & \colhead{DM$^{b}$} & \colhead{Duration$^{b}$} & \colhead{Component \#$^{b}$} & \colhead{Width$^{b}$} & \colhead{$\gamma$$^{b}$} & \colhead{$\beta$$^{b}$} & \colhead{$\Pi_L$} & \colhead{$\rm{RM}$} \\
\colhead{} & \colhead{} & \colhead{(Jy ms)} & \colhead{(pc cm$^{-3}$)} & \colhead{(ms)} & \colhead{} & \colhead{(ms)} & \colhead{} & \colhead{} & \colhead{} & \colhead{($\rm{rad\, m^{-2}}$)}
}
\startdata
FRB 20210601C & 111.1 & 30(3) & 419.55(1) & 9.58(7) & 1 & 4.07(4) & $-$ & $-$ & 0.69(2) & $-110.0(2)$ \\
\hline
\multirow{4}{*}{FRB 20210601F} & \multirow{4}{*}{62.6} & \multirow{4}{*}{17(2)} & \multirow{4}{*}{418.96(2)} & \multirow{4}{*}{13.75(7)} & 1 & 0.97(6) & 10(1) & $-$39(6) & \multirow{4}{*}{0.80(3)}& \multirow{4}{*}{$-111.4(4)$} \\
 & & & & & 2 & 5.0(4) & 1.5(3) & 6(1) & & \\
 & & & & & 3 & 0.32(2) & 48(5) & $-$126(10) & & \\
 & & & & & 4 & 0.68(4) & 16(2) & $-$81(10) & & \\
\enddata
\tablecomments{
\\
$^a$ Values in parentheses indicate $1\sigma$ uncertainties on the last quoted digit(s). Baseband-determined TOAs are omitted owing to the uncertainty in the absolute timing offset between the CHIME/FRB baseband and intensity data systems. See Table~\ref{ta:summary} for the intensity-determined topocentric TOAs.\\
$^b$ {\tt fitburst} determined parameters. \\
}
\end{deluxetable*}
\clearpage

\begin{figure}
	\centering
\includegraphics[width=0.52\textwidth]{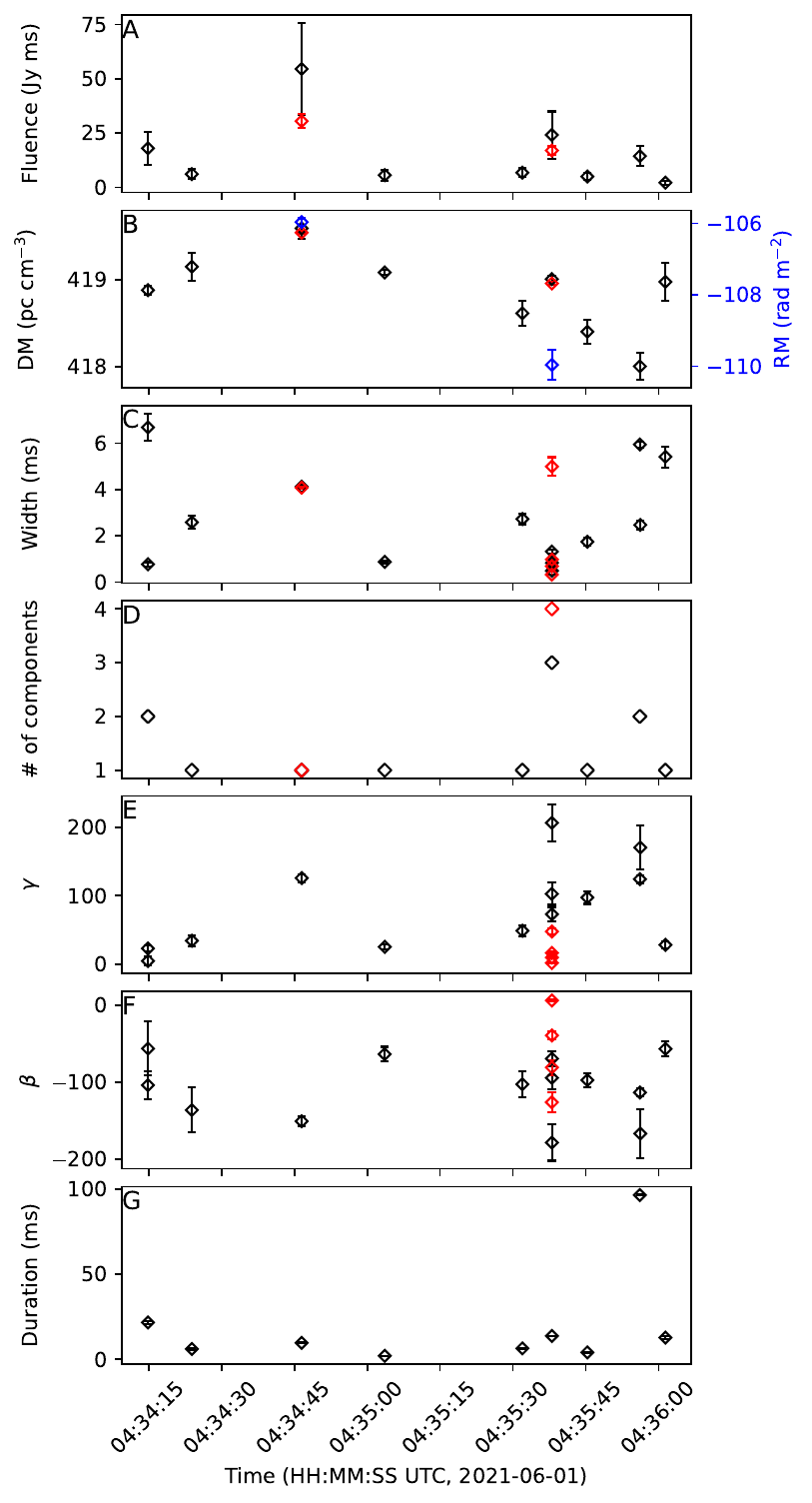}
\caption{Burst properties (Fluence, DM, subcomponent width, number of components, $\gamma$, $\beta$, burst envelope duration) versus time for FRB 20210601A (Table~\ref{ta:summary}). Red symbols indicate equivalent measurements determined from baseband data for repeat bursts FRB 20210601C \& FRB 20210601F (Table~\ref{ta:summary_baseband}). Panel B includes an additional twin y-axis capturing RM measurements (blue) from these two bursts. Plotted uncertainties correspond to measurement error and do not account for additional systematic error suspected to be present for certain properties (e.g., DM, RM, $\gamma$, $\beta$; see text for details).}
\label{fig:prop_vs_time}
\end{figure}

\clearpage

\begin{figure*}
	\centering
\includegraphics[width=0.43\textwidth]{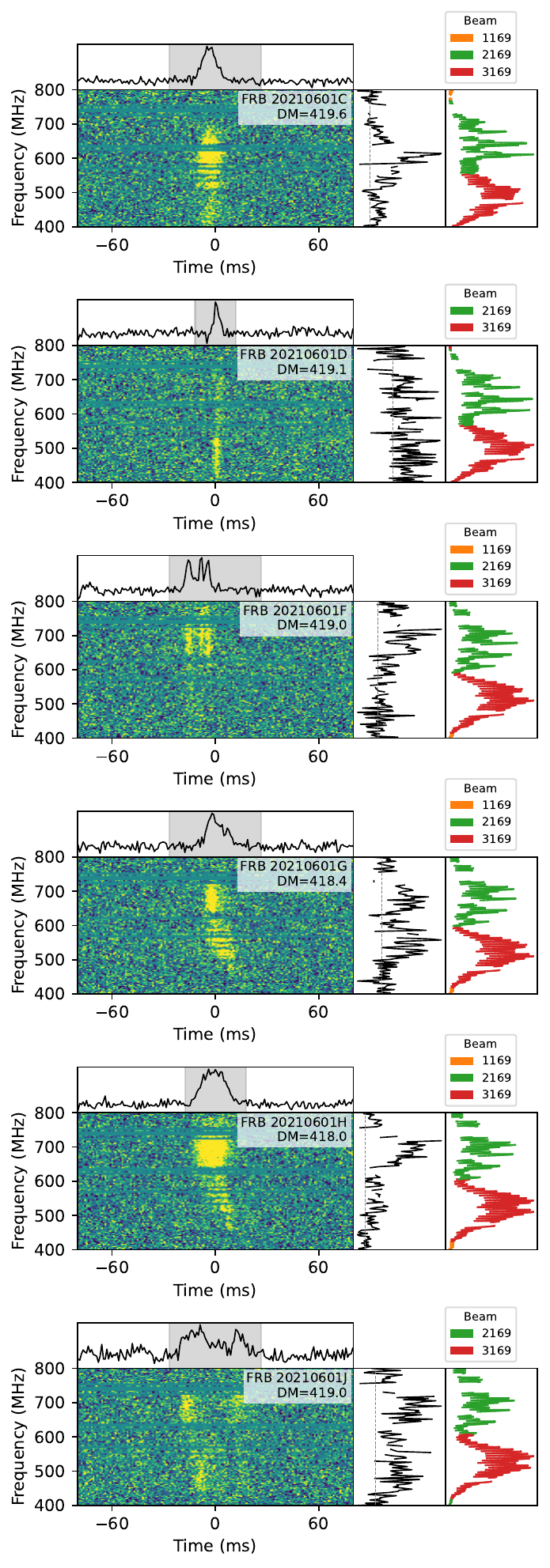}
\includegraphics[width=0.43\textwidth]{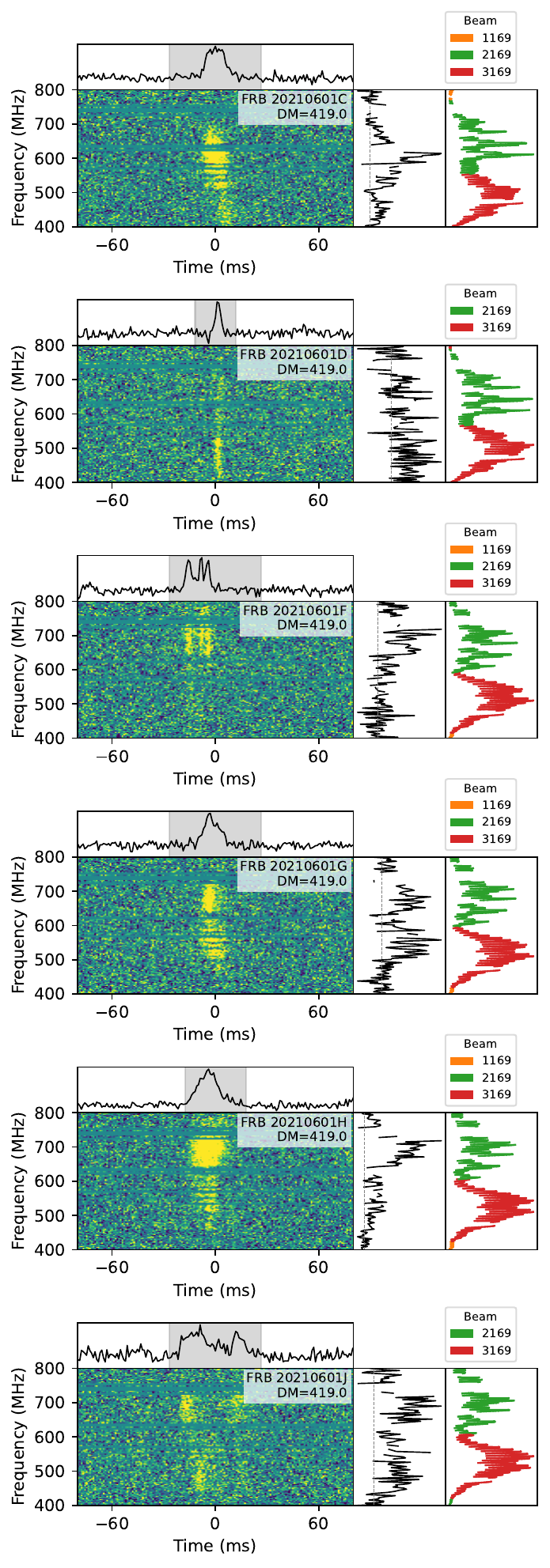}
\caption{Composite beam waterfall plots for the six bursts from FRB~20210601A with intensity data available from adjacent CHIME/FRB FFT beams. Each row shows one burst dedispersed to its {\tt fitburst}-determined DM (\textit{left}) and to the weighted average $\mathrm{DM_{avg}} = 419.01 \pm 0.02~\mathrm{pc~cm^{-3}}$ (\textit{right}). Each panel shows the frequency-averaged time profile (top) and sensitivity-weighted composite waterfall (bottom), with each frequency channel drawn from the FFT beam of highest predicted sensitivity. Side panels show the burst spectrum, averaged over the gray-shaded window (\textit{centre}), and the maximum beam sensitivity versus frequency, colour-coded by contributing beam (\textit{right}; see legend).}
\label{fig:composite_waterfall}
\end{figure*}

\begin{figure*}
	\centering
\includegraphics[width=0.4\textwidth]{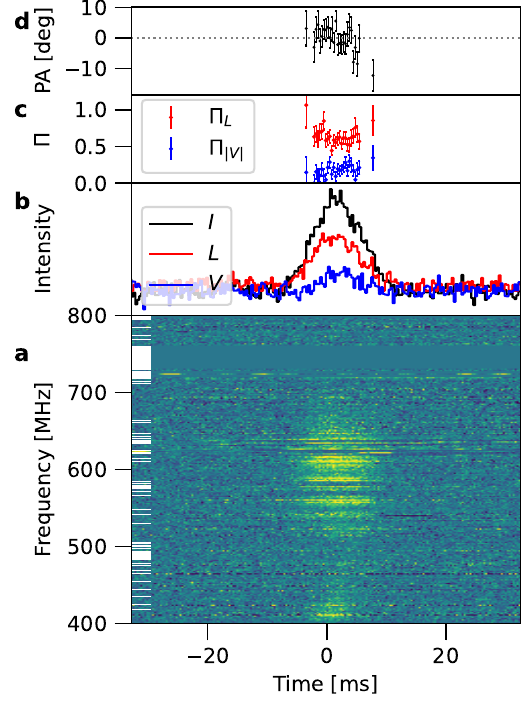} 
\includegraphics[width=0.4\textwidth]{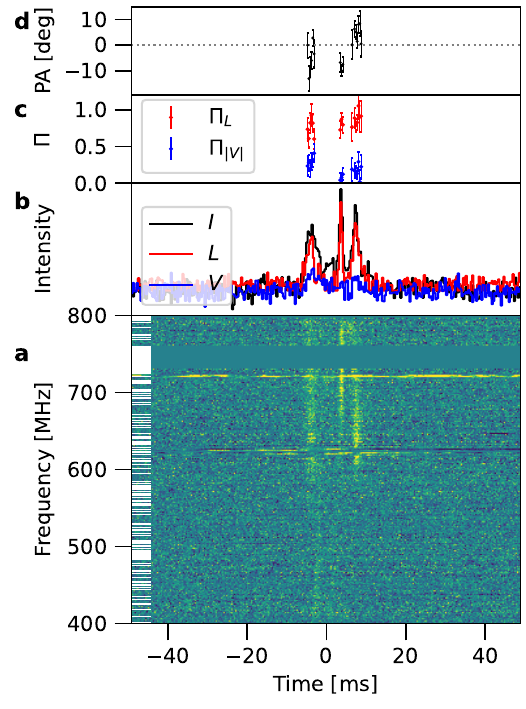} 
\caption{High temporal resolution waterfalls plots and associated profiles of FRB 20210601C (left panel) and FRB 20210601F (right panel) determined from CHIME/FRB voltage (baseband) data. \textit{Panel a:} Dynamic spectra dedispersed to their individual baseband-determined {\tt{fitburst}} DMs (see Table~\ref{ta:summary_baseband}). Data have been rebinned from their native 0.390625 MHz/bin and $2.56\; \rm{\mu s}$/bin resolutions to 1.5625 MHz/bin and  0.32768 ms/bin. Masked frequency channels are indicated by white lines along the vertical axis. \textit{Panel b:} Burst profiles of total (black), linear polarized (red), and circularly polarized (blue) intensity (peak normalized). \textit{Panel c:} Polarization fractions profiles for the linear (red) and circularly (blue) polarized signal. \textit{Panel d:} The polarization angle (PA) profile. Burst profiles are obtained by adding signal over the spectral limits of the burst.}
\label{fig:baseband_waterfall}
\end{figure*}

\clearpage

\begin{figure*}
	\centering
\includegraphics[width=0.8\textwidth]{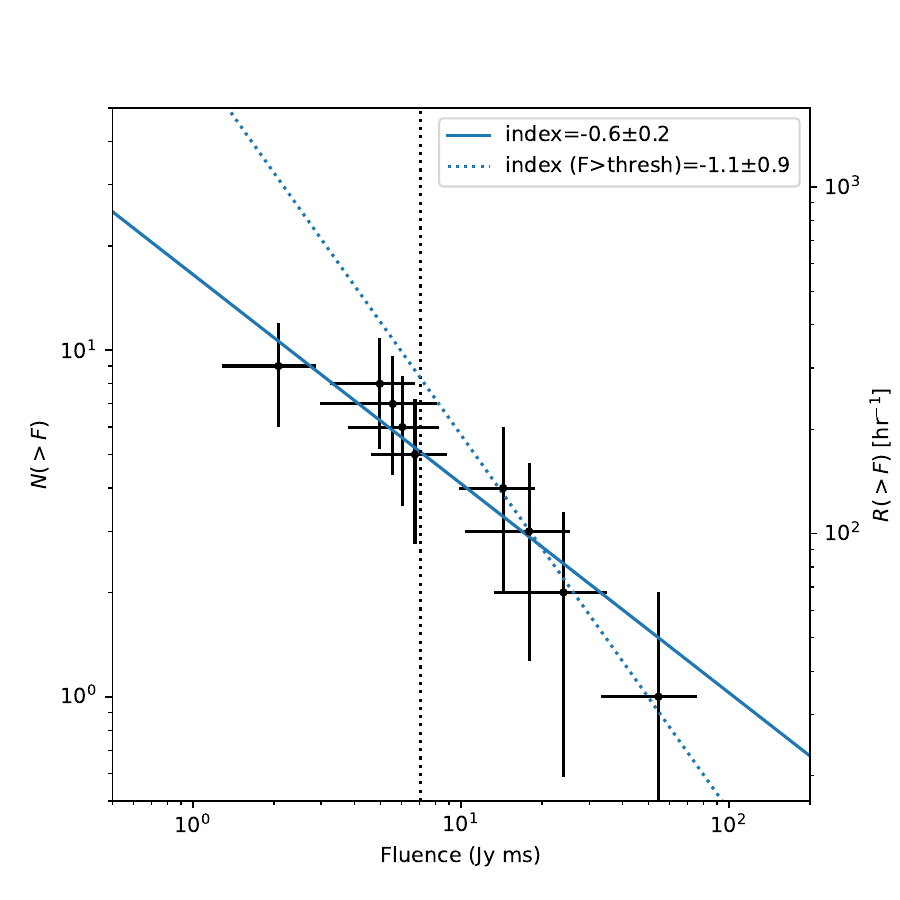}
\caption{Cumulative fluence distribution of the nine bursts detected from FRB 20210601A during its single active transit. The cumulative number of bursts with fluence greater than $F$, $N(>F)$, is shown as a function of fluence, with Poisson uncertainties on the counts. The 95\% fluence threshold estimate for the source (7 Jy ms) is indicated by the vertical dotted line and is revised from value reported for the source in CHIME/FRB Catalog 2 using the updated baseband position. The blue solid line shows the best-fit power-law model $N(>F)\propto F^{\rm{index}}$, yielding a cumulative fluence index = -0.6(0.2), while the blue dotted line shows the equivalent fit restricted to bursts above the fluence threshold, yielding an index of -1.1(0.9).}
\label{fig:fluence_dist}
\end{figure*}

\end{document}